%% file: main.tex
\documentclass[
preprint, 
superscriptaddress, amsmath, amssymb, aps, pre]{revtex4}
\usepackage{lineno}
\usepackage{graphicx}
\usepackage{dcolumn}
\usepackage{bm}
\usepackage{booktabs}
\usepackage{subcaption}
\usepackage{hyperref}
\usepackage{makecell} 
\usepackage{threeparttable} 
\usepackage{amsmath}

\begin{document}


\title{Automated structure discovery for Tip Enhanced Raman Spectroscopy}

\author{Harshit Sethi} 
\affiliation{Department of Applied Physics, Aalto University, Helsinki, FI-02150, Finland}
\author{Markus Junttila}
\affiliation{Department of Applied Physics, Aalto University, Helsinki, FI-02150, Finland}
\author{Orlando J. Silveira}
\email{orlando.silveirajunior@aalto.fi}
\affiliation{Department of Applied Physics, Aalto University, Helsinki, FI-02150, Finland}
\author{Adam S. Foster}
\email{adam.foster@aalto.fi}
\affiliation{Department of Applied Physics, Aalto University, Helsinki, FI-02150, Finland}
\affiliation{WPI Nano Life Science Institute (WPI-NanoLSI), Kanazawa University, Kakuma-machi, Kanazawa, 920-1192, Japan}

\begin{abstract}
Tip‑Enhanced Raman Spectroscopy (TERS) provides nanoscale chemical fingerprints alongside high‑resolution topographic mapping of molecules, offering a powerful tool for materials discovery. However, TERS image datasets are challenging to interpret and typically demand time‑consuming, computationally intensive quantum‑chemistry calculations. To overcome this problem, we present an encoder–decoder model trained and evaluated on simulated TERS images of planar molecules, enabling direct prediction of molecular structures from spectral simulated data with high accuracy. Our approach demonstrates the feasibility of automating molecular structure identification from TERS images, bypassing traditional manual analysis. These findings provide a foundation for extending machine learning methods to experimental TERS datasets, potentially accelerating molecular discovery by integrating nanoscale spectroscopy with automated computational analysis. 
\end{abstract}

\maketitle
\section{Introduction}
\label{sec:intro}

\begin{figure}
    \centering
    \includegraphics[width=0.7\linewidth]{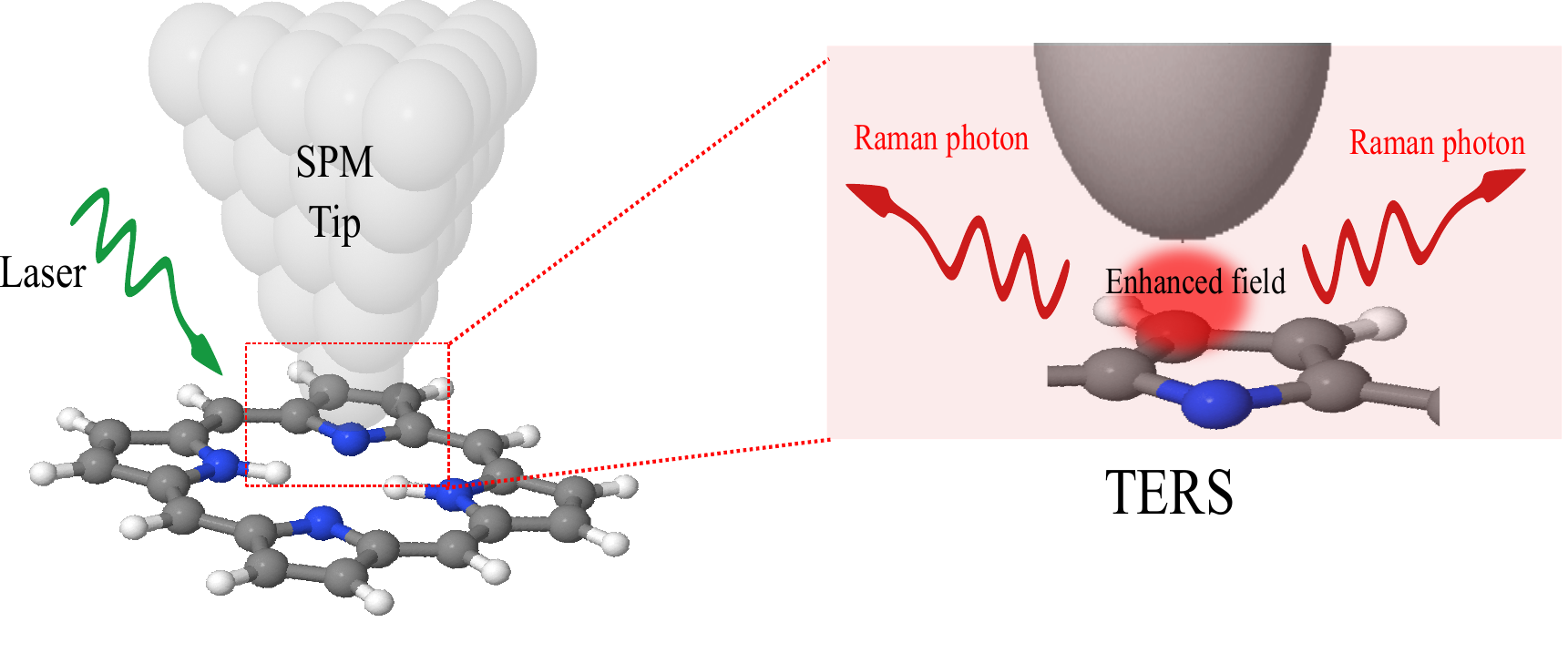}
    \caption{Schematic illustration of the tip-enhanced Raman spectroscopy (TERS) setup, where laser illumination induces localized field enhancement at the SPM tip apex, enabling enhanced Raman signal intensity.}
    \label{fig:ters_schematic}
\end{figure}

Achieving atomic resolution in optical imaging is a central goal for advancing nanoscale science and enabling the study and design of materials. In Scanning Tunneling Microscopy (STM), the functionalization of a tip apex with a single CO molecule enables unprecedented resolution of the internal structure of adsorbed molecules and their orbital densities \cite{mohn2010,prokop2014,Sun2023,niko2024}. Similar functionalization in non-contact Atomic Force Microscopy (AFM) provides stunning sub-atomic resolution of the skeletal structure of a molecule \cite{gross_chemical_2009,gross2012,prokop2014,niko2024}. Both STM and AFM are among a family of techniques called scanning probe microscopy (SPM), and while it has been proposed that combining different modes of SPM allows for resolution not only the geometric arrangement of the atoms, but also their chemical composition, formidable challenges are still faced to achieve this routinely \cite{Schulz2018}. In contrast, conventional chemically specific non-destructive spectroscopies, such as infrared and Raman, readily offer chemical characterization, but are constrained by the Abbe diffraction limit and thus lack the type of spatial resolution that CO-tip functionalized SPM can give \cite{Abbe1881, Itoh2023}.  However, recent advances have shown that integrating an STM tip to the Raman spectrometer greatly enhances the Raman signal, giving the possibility to render both topographical and chemical information by imaging molecules at the Ångström scale \cite{zrimsek2017,yang2023,Itoh2023,Lee2019,Jiang2017,kumagai2025,Ferreira_experimental}. In this integrated technique, named tip enhanced Raman spectroscopy (TERS), an incoming light triggers a collective response to the electrons at the STM tip apex, producing localized surface plasmon excitations and confining propagating light \cite{hoppener2024} (see Fig. \ref{fig:ters_schematic}). The STM tip acts then as a nanoantenna, transferring to the far-field information from the evanescent waves that originate from interactions of the light with sample sizes smaller than its wavelength, thus overcoming the diffraction limit and limiting the spatial resolution only to the apex radius.
Despite these advances, the interpretation of TERS images remains challenging, owing to complex contrast mechanisms, tip dependence, and the large volume and dimensionality of the resulting datasets. Theoretical developments have successfully allowed individual comparisons of simulated TERS images and experiments through visual inspections \cite{Lee2019,Zhang2021,Duan2015,zhang_matlab_2023,litman2023,Chen2019,Qiu2022,Yu2026,brezina2025}, but they can be quite demanding tasks as TERS grows to belong to the mainstream group of experiments for molecular characterization \cite{chen_ml_optical,Kumar2026}. All these challenges naturally motivate the use of data-driven approaches to assist in extracting physically and chemically meaningful information from such measurements. 



Machine learning (ML) methods have been increasingly used in microscopy to assist with the analysis and interpretation of large, noisy, and high-dimensional datasets. Applications include image denoising and reconstruction, feature identification, and the extraction of underlying physical trends that are difficult to access with conventional analysis. 
In scanning transmission electron microscopy (STEM), for example,
CNNs have been applied to STEM images to extract structural and compositional information at high throughput \cite{osti_1376377}. Domain-adaptation methods such as CycleGANs have been employed to render simulated STEM data more realistic for training, enabling robust identification of single-atom defects \cite{khan2023}. More recently, vision-language transformers have been used to predict atomic structure as captions from STEM images \cite{chaudhary_microscopy_gpt}. For SPM in particular, recent advances at the intersection with ML have advanced prospects for autonomous materials discovery \cite{kalinin_big_2016,rahman_laskar_scanning_2023}. Early supervised ML studies established that SPM images can be mapped directly to atomic-scale structures. For example, by using  convolutional neural networks (CNNs) to directly predict likely 2D maps of structures from AFM images \cite{ben2020,tang_machine_2023}, or combined with graph neural networks (GNNs) \cite{Oinonen2022} for both STM and AFM \cite{Kurki2024,fabio2024}. Complementary approaches used multimodal recurrent networks to infer IUPAC names from AFM data \cite{carr_iupac}, and conditional generative adversarial networks (cGANs) to translate AFM images into ball-and-stick representations \cite{carracedocosme2024}. 

In this work, we present \textbf{SMARTERS} (Structural Machine Learning for High-Resolution Tip-Enhanced Raman Spectroscopy), an automated CNN-based ML model motivated by the growing use of ML in microscopy and the need for theoretical tools to decode TERS images.
Inspired by the success of SPM models trained on simulations, here, the dataset of TERS images was simulated by considering the plasmonic resonance created at the SPM tip as the main source of enhancement of the Raman signal.
Within this approach, the local plasmonic field is then treated as a three-dimensional Gaussian function with a specific full width at half maximum (FWHM) \cite{Duan2015,Chen2019,duan2016,li2025,Qiu2022,Yu2026}.
Despite the local field created on the tip, it is well known that the interactions of the molecules with the underlying substrate also play an important role in the enhancement of the Raman signal \cite{litman2023,brezina2025}, and subtle variations of adsorption geometries may drastically affect the vibrations of the molecules \cite{Ferreira_experimental}. In our simulations, we considered only isolated molecules, making the model more suitable for planar molecules lying flat on top of weakly interacting substrates, e.g.\ NaCl, where disturbances in the vibrational modes are expected to be less \cite{yang2023}. Additionally, since different molecules exhibit varying numbers of vibrational modes, we proposed a binning procedure along the frequency axes to uniformize the hyperspectral image with different numbers of spectral channels, making the data more suitable for the neural network training.

SMARTERS  provides a strong methodological foundation for automated scanning probe microscopy platforms and represents a step toward realizing similar autonomous capabilities in TERS, possibly enabling in the future chemically specific discovery and characterization at the atomic and molecular scale.
Automated molecular structure discovery from TERS hyperspectral images has not, to our knowledge, been demonstrated prior to this work. Here, we address that gap by developing and validating ML methods for structural discovery from TERS hyperspectral datasets. Our model serves as a proof of concept for automating molecular structure identification from TERS images and contributes to a better understanding of light-matter interactions.

\section{Methods}
\label{sec:methods}

\begin{figure}
    \centering
    \includegraphics[width=0.8\linewidth]{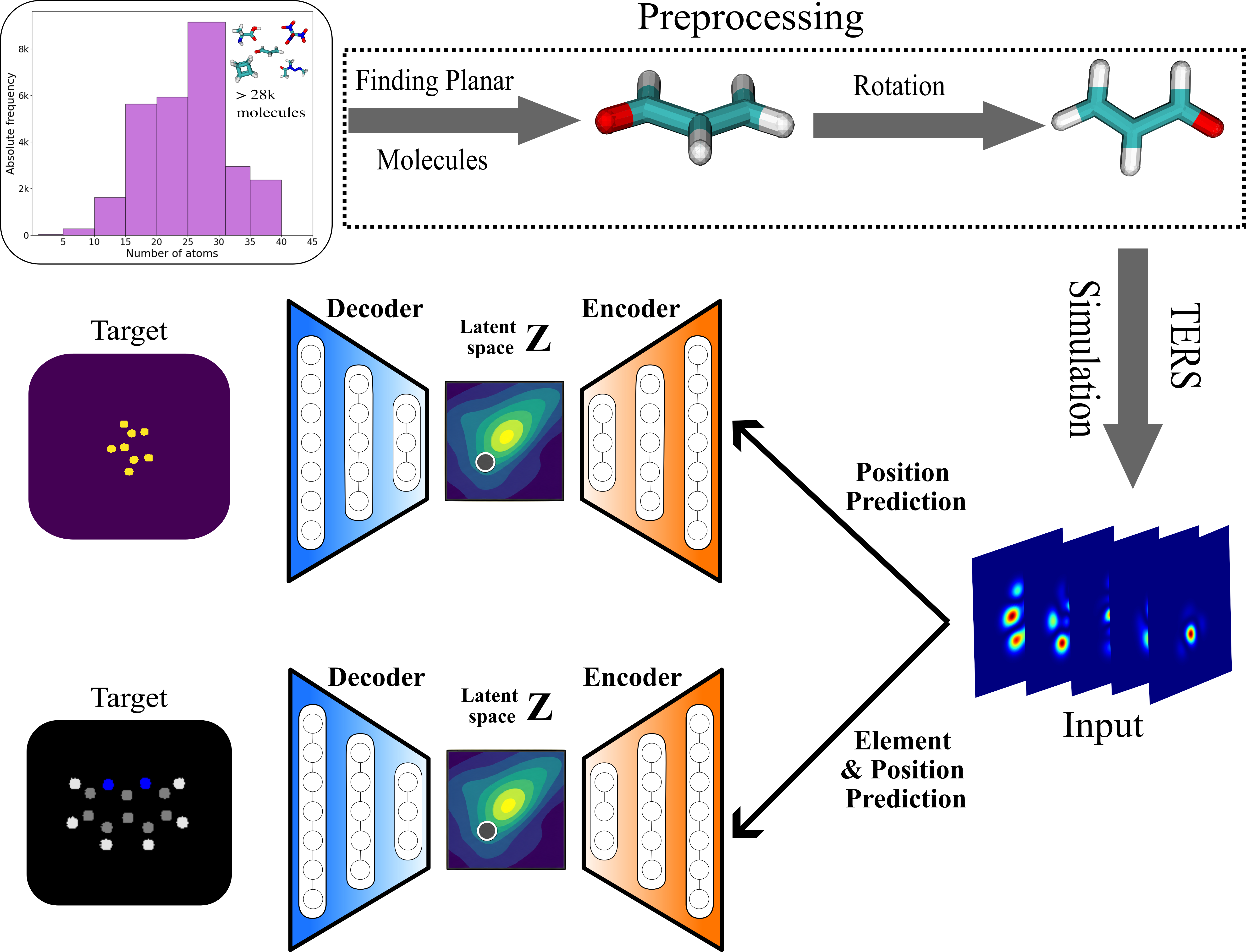}
    \caption{Schematic of the deep learning workflow for structure prediction from TERS images.}
    \label{fig:pipeline}
\end{figure}

We formulate the task of determining the geometry of molecules from TERS data as an image-to-image translation problem. To address this, the methodology used in SMARTERS is centered on training a deep encoder-decoder network on a large-scale, computationally simulated dataset. This process involves curating a cohort of suitable planar molecules, then for each one, generating a synthetic TERS spectral image paired with its ground-truth 2D atomic position map. The TERS images are generated using our Python re-implementation of the simulation method first implemented by Zhang et al.~\cite{zhang_matlab_2023}. The network is then trained in a supervised manner to learn a direct mapping from the TERS image to the atomic map, where pixel intensity represents the presence of an atom. The entire end-to-end pipeline is depicted in Fig. \ref{fig:pipeline}.

\subsection{Molecule Dataset}

\begin{figure}[h!]
    \centering

    \begin{subfigure}[b]{0.4\textwidth}
        \centering
        \includegraphics[width=\textwidth]{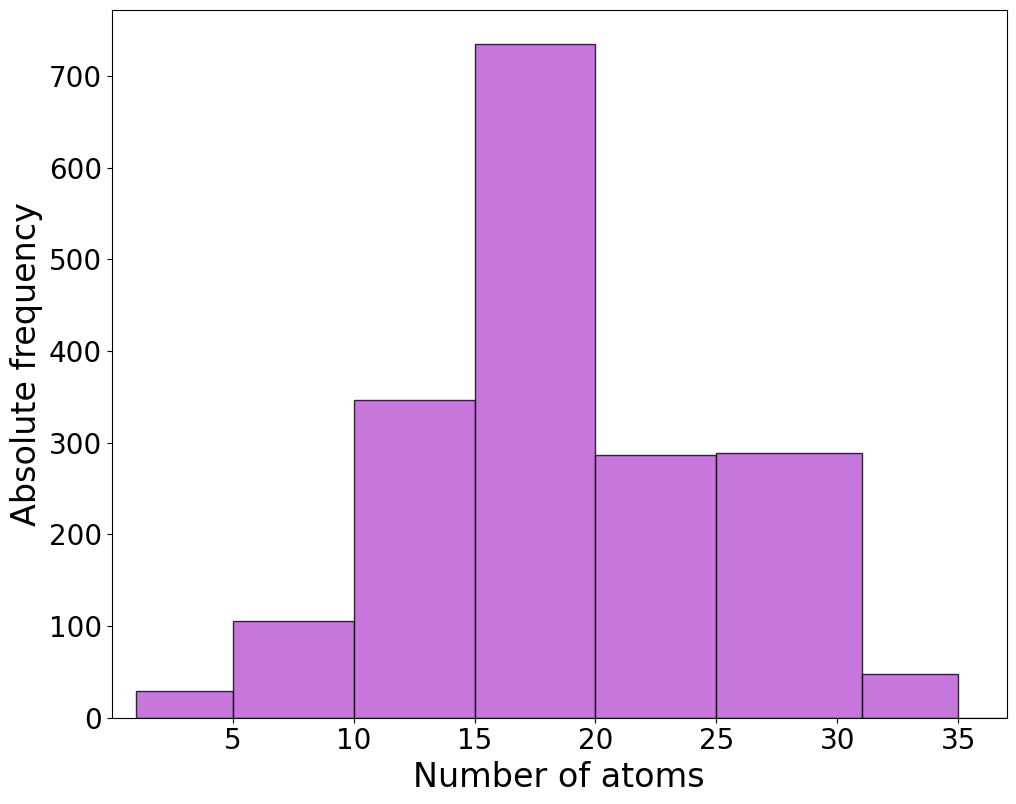}
        \caption{Distribution of molecule size (number of atoms).}
    \end{subfigure}
    \hspace{0.02\textwidth}
    \begin{subfigure}[b]{0.4\textwidth}
        \centering
        \includegraphics[width=\textwidth]{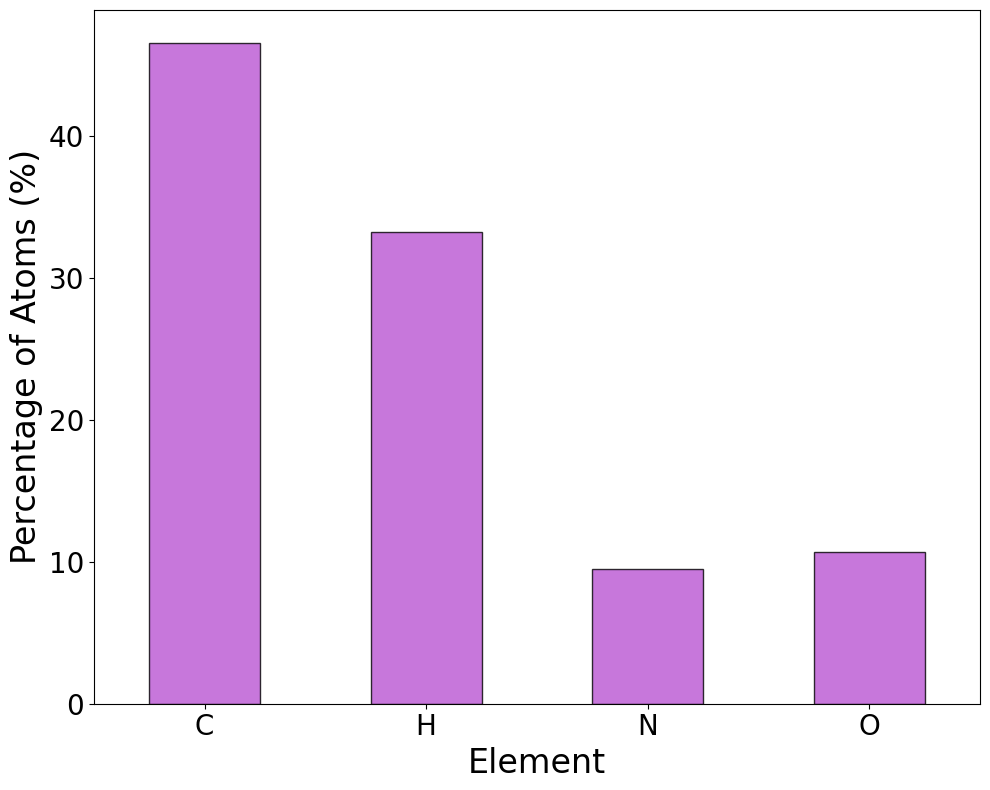}
        \caption{Distribution of constituent elements.}
    \end{subfigure}

    \caption{Characteristics of the final dataset of 1,840 preprocessed planar molecules used for simulated TERS image generation.}
    \label{fig:mol_dataset}
\end{figure}

For generating TERS simulated images, the geometries of sample molecules were taken from a well-known database of small organic molecules structurally optimized with density functional theory (DFT) \cite{Oinonen_mol_dataset}. The selected dataset was a sub-part of the original dataset chosen for establishing automating structure discovery in AFM -- we initially selected a subset of 28,570 molecules containing up to 40 atoms as representative of commonly studied systems in TERS. From this subset, we systematically identified planar molecules by applying principal component analysis (PCA) to the atomic coordinates of each molecule, retaining a molecule when its root-mean-square out-of-plane deviation was at most 0.05~\AA\. However, planarity alone was insufficient to ensure optimal TERS imaging conditions as planar molecules can be arbitrarily rotated, potentially obscuring parts of the structure from the tip's view during a scan. We mitigated this by rotating each selected molecule to lie in the plane, perpendicular to the tip's axis (see Appendix \ref{sec:preprocessing} for details). Following this preprocessing pipeline, the dataset was reduced to 1840 molecules, which were subsequently used to generate TERS hyperspectral images for model training. The statistical properties of this set, including distributions of molecule size and elemental composition, are presented in Fig. \ref{fig:mol_dataset}.

\subsection{Simulation Method}
\label{subsec:simulation_method}

\begin{figure}[h!]
    \centering
    \includegraphics[width=0.99\linewidth]{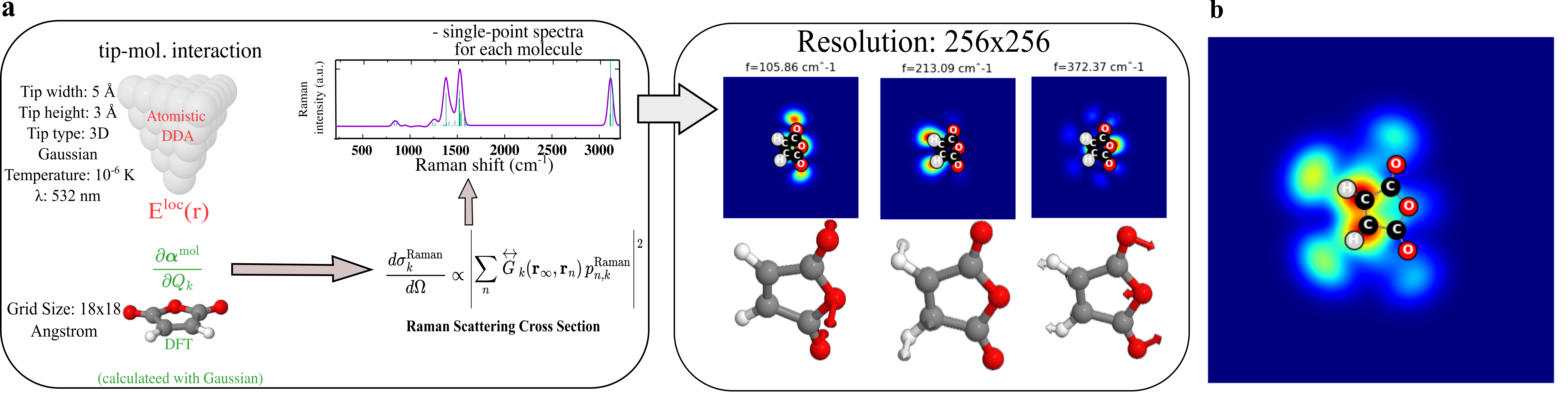}

    \vspace{-0.5em}
    
    \caption{\textbf{TERS Simulation Pipeline.} Our data generation process, adapted from Zhang et al.~\cite{Zhang2021}. (a) The workflow takes a molecule's geometry, calculates DFT properties, simulates interaction, and computes the Raman spectrum. (b) Sum of all TERS Hyperspectral images for visualization.}
    \label{fig:sim_pipeline}
\end{figure}

Our simulation approach to calculate the intensity of the Raman scattering is based on previous work \cite{Duan2015,Chen2019,duan2016,li2025,Qiu2022,Yu2026,Zhang2021}, while the complete workflow of the simulation method is illustrated in Fig. \ref{fig:sim_pipeline}. In SMARTERS, we ported the code developed by Yao Zhang \textit{et al.} \cite{Zhang2021} into Python to ensure compatibility within our ML framework. The simulation technique decouples the calculation of contribution of the tip and the molecule.
In this approach, the induced dipole moment $\boldsymbol{\mu}$ of a molecule polarized by an electric field $\boldsymbol{E}$ can be expanded in terms of the normal coordinates $Q_k$ of the $k$th vibrational mode
\begin{equation}
        \boldsymbol{\mu} = \alpha \boldsymbol{E} = \alpha_0 \boldsymbol{E} + \sum_k  \left ( \frac{\partial \alpha}{\partial Q_k} \right )_0 Q_k \boldsymbol{E} + \cdots ,
        \label{eq1}
\end{equation}
where $\alpha$ is the polarizability of the molecule. The Raman scattering process of the molecule is given by the second term of Eq.~\eqref{eq1} in the form of a Raman dipole moment. By applying a transformation from normal coordinates $Q_k$ to atomic coordinates $r^{(n)}$ in the derivative of the polarizability, one can write
\begin{equation}
    \frac{\partial \alpha}{\partial Q_k} = C \sum_{n = 1}^{M} L_k^{(n)} \frac{\partial \alpha}{\partial r^{(n)}} = \sum_{n=1}^M \alpha_k^{(n)} ,
\end{equation}
where $C$ is a constant, $M$ is the number of atoms in the molecule, $L_k^{(n)}$ is the normalized displacement of the $n$th atom corresponding to the vibrational mode $k$, and $r^{(n)}$ is the coordinate of the $n$th atom. The $\alpha_k^{(n)}$ is defined as an atomistic polarizability of the atom $n$ corresponding to vibrational mode $k$, allowing the understanding of the contribution of each atom to the Raman scattering. The atomistic dipole moment due to an incident inhomogeneous local field induced by the STM tip will then become
\begin{equation}
    \boldsymbol{\mu}_n^k = \alpha_k^{(n)} \boldsymbol{E}^{loc} (\boldsymbol{r}_n),
\end{equation}
and ultimately the Raman scattering cross section will be given by adding the contribution from each atom in the molecule \cite{Zhang2021}. The effect induced by the localized field is to enhance the atomistic Raman dipole moment, as well as the radiation of the molecular dipole. Finally, simulated TERS hyperspectral images corresponding to different vibrational modes are generated by combining spectral information from various spatial regions of the molecule. In our implementation, polarizabilities and their derivatives were calculated using DFT with the Gaussian software package \cite{g16}.

One of the most important parameters for the TERS simulated images is the size of the spatially confined plasmon, which is implicitly defined by the incident inhomogeneous local field $\boldsymbol{E}^{loc} (\boldsymbol{r}_n)$. The more localized the plasmon, the more subtle features of the spatial distribution of the vibrational modes can be identified, thus enabling the model to better distinguish between structurally similar molecules. In contrast, a larger plasmon is not able to distinguish between important spectral features, reducing the prediction reliability of our model with respect to those features. Our images were then generated by setting a plasmonic size of 5 Å (represented by a 3D gaussian with FWHM of 5 Å), which was enough to distinguish, for example, features arising from Jahn–Teller distortions \cite{Yu2026}. The simulation field of view was set to $18 \times 18~\text{Å}^2$ so that the full lateral extent of the largest molecules is contained within the scan window. Smaller grids would truncate peripheral regions of the structure and effectively crop the simulated images. The hyperspectral maps were rasterised at $256 \times 256$ pixels to preserve spatial detail. Coarser discretisations reduced performance by limiting the model’s ability to distinguish vibrational mode signals from noise. 

\subsection{Image Dataset}
\label{sec:image_dataset}

Using the simulation method described above, we generated TERS hyperspectral image cubes for each of the 1,840 planar molecules in our dataset. A key challenge arose from the fact that different molecules exhibit varying numbers of vibrational modes, resulting in hyperspectral images with different numbers of spectral channels. To obtain a uniform representation suitable for neural network training, we developed a binning procedure along the frequency axis. The binning strategy divides the frequency range of interest (\(0\)–\(4000 \,\text{cm}^{-1}\)) into \(n\) uniform bins, where the \(i\)-th bin covers the interval $\Big[i \cdot \tfrac{4000}{n}, \, (i+1) \cdot \tfrac{4000}{n}\Big], 
\quad i = 0, 1, \dots, n-1 $. Each bin aggregates vibrational contributions within its frequency range, resulting in a standardized hyperspectral cube of size \(H \times W \times n\), where \(H\) and \(W\) denote the spatial dimensions, and \(n\) represents the number of spectral bins (channels). This binning procedure provides two main advantages: 1) it produces hyperspectral cubes with a fixed number of channels across all molecules, enabling consistent model training; 2) it preserves coarse frequency localization of vibrational modes while reducing computational complexity.

We treated $n$ as a hyperparameter and systematically compared different settings with $n = 400$ and $n = 100$. The region of interest for frequency was from $0$-$4000~\text{cm}^{-1}$. Therefore, $n = 400$ corresponds to a spectral resolution of $10~\text{cm}^{-1}$ per bin, while $n = 100$ corresponds to $40~\text{cm}^{-1}$ per bin. Based on validation-set performance (Sec.~\ref{sec:evaluation}), we determined that $n = 100$ provided better accuracy while simultaneously reducing training time due to the smaller input dimensionality.

\subsection{Model training}
\label{subsec:model_training}

The ML architecture of SMARTERS is similar to the encoder-decoder–type networks that have widely been used in computer vision problems, for example, image segmentation \cite{badri2017segnet}. We selected the Attention U-Net architecture \cite{oktay2018attention} as the backbone for both of our prediction tasks. The same architecture is used in two configurations that differ only in the number of output channels: a \emph{atomic structure prediction} model that outputs a single-channel atomic map, where pixel intensity encodes the presence of an atom irrespective of its species; and a \emph{chemically-resolved structure prediction} model that outputs a four-channel map, one channel per element (H, C, N, O), encoding both atomic position and elemental identity.
We chose this model motivated by its demonstrated prior success in automating structure discovery from scanning probe microscopy images \cite{ben2020, Kurki2024, Oinonen2022}. The number of input channels of the model was treated as a hyperparameter to accommodate different spectral representations of the simulated TERS hyperspectral images as input, with either 100 or 400 channels depending on the binning strategy described in Section \ref{sec:image_dataset}. This design allowed us to assess whether higher spectral resolution provides additional predictive benefit, or whether a reduced representation obtained by summing nearby frequency bands is sufficient.

The ML model operates through two main phases: feature extraction and structure prediction. In the encoder phase, the network extracts spectroscopic features from each frequency channel as a function of their spectral character and spatial position. This process occurs simultaneously across all frequency channels in the hyperspectral data. Spectroscopic features that appear consistently across multiple frequency channels yield stronger aggregated activations in the convolutional layers. As the network processes data through its multiple layers, it filters these spectroscopic features according to learned weights and biases. This mathematically prioritizes persistent multi-channel signals over uncorrelated background noise, ultimately generating critical feature maps that serve as statistical proxy for molecular structural elements. The decoder then reconstructs the 2D atomic map from these learned features, making predictions about atomic positions based on the spectroscopic input. The attention mechanism in the U-Net architecture is particularly valuable for TERS data analysis, as it allows the network to focus on the most informative spectral-spatial relationships while suppressing irrelevant background signals. This is crucial for TERS applications where signal-to-noise ratios can vary and specific vibrational modes may provide more structural information than others.

Our dataset of 1,840 planar molecules was split into 80/10/10 partitions for training, validation, and testing respectively. The network was trained for 50 epochs using the Adam optimizer \cite{kingma2017adam}. 
Performance is quantified by the mean Dice Similarity Coefficient (DSC) \cite{dice_measures_1945}, which measures the pixel-wise overlap between the predicted and ground-truth maps. Although DSC is a region-overlap metric, in our sparse-map setting it is strongly sensitive to atom position, so misplaced atoms are penalized rather than tolerated (see Appendix~\ref{sec:dice_score} for more details). For the atomic structure prediction model the DSC is computed over the single foreground channel, whereas for the chemical species prediction model it is computed per element and averaged across the four channels, giving each species equal weight regardless of its abundance in the dataset.
DSC loss is well suited to both prediction tasks, as it directly optimizes for spatial overlap and is robust to class imbalance. For the structure prediction model, this addresses the imbalance between sparse atomic pixels and the background. For the chemically-resolved structure prediction model, the loss is computed per channel and averaged across the four elements, which additionally mitigates the imbalance arising from the underrepresentation of nitrogen and oxygen relative to carbon and hydrogen.
Simple data augmentations were applied on-the-fly during training to encourage invariance to image orientation and robustness to additive noise, serving as a deliberately simplified proxy for common experimental variations. The augmentation pipeline included random rotations by multiples of $90^\circ$, horizontal and vertical flips, and additive Gaussian noise with standard deviation sampled uniformly from $[0.01,\,0.1]$. Identical geometric transformations were applied to the input images and corresponding masks. Validation and test sets were kept unchanged.

To optimize model performance, we conducted a systematic exploration of key hyperparameters including batch size, learning rate, encoder/decoder depth, and the number of attention channels. We employed Bayesian optimization using the Optuna library \cite{akiba2019optuna} to efficiently navigate the hyperparameter space. The explored ranges and final selected values are summarized in Table~\ref{tab:hyperparams} in Appendix \ref{sec:model}. Our entire implementation is built on the PyTorch framework~\cite{paszke2019pytorch}.

\section{Results}

\subsection{Atomic position prediction}
\label{sec:evaluation}
We evaluate our proposed Attention U-Net model on the task of atomic position prediction for planar molecules. The model is trained to generate a labeled atom map from simulated images, where each atom is represented as a circle of fixed radius. Our model achieves a mean DSC of 0.862, 0.833 and 0.842 on the training, validation and test set respectively, demonstrating its efficacy in predicting the geometry of the molecule. The performance across the training, validation, and test splits remains consistent, indicating robust generalization.

\begin{figure}[h!]
    \centering
    \includegraphics[width=0.9\linewidth]{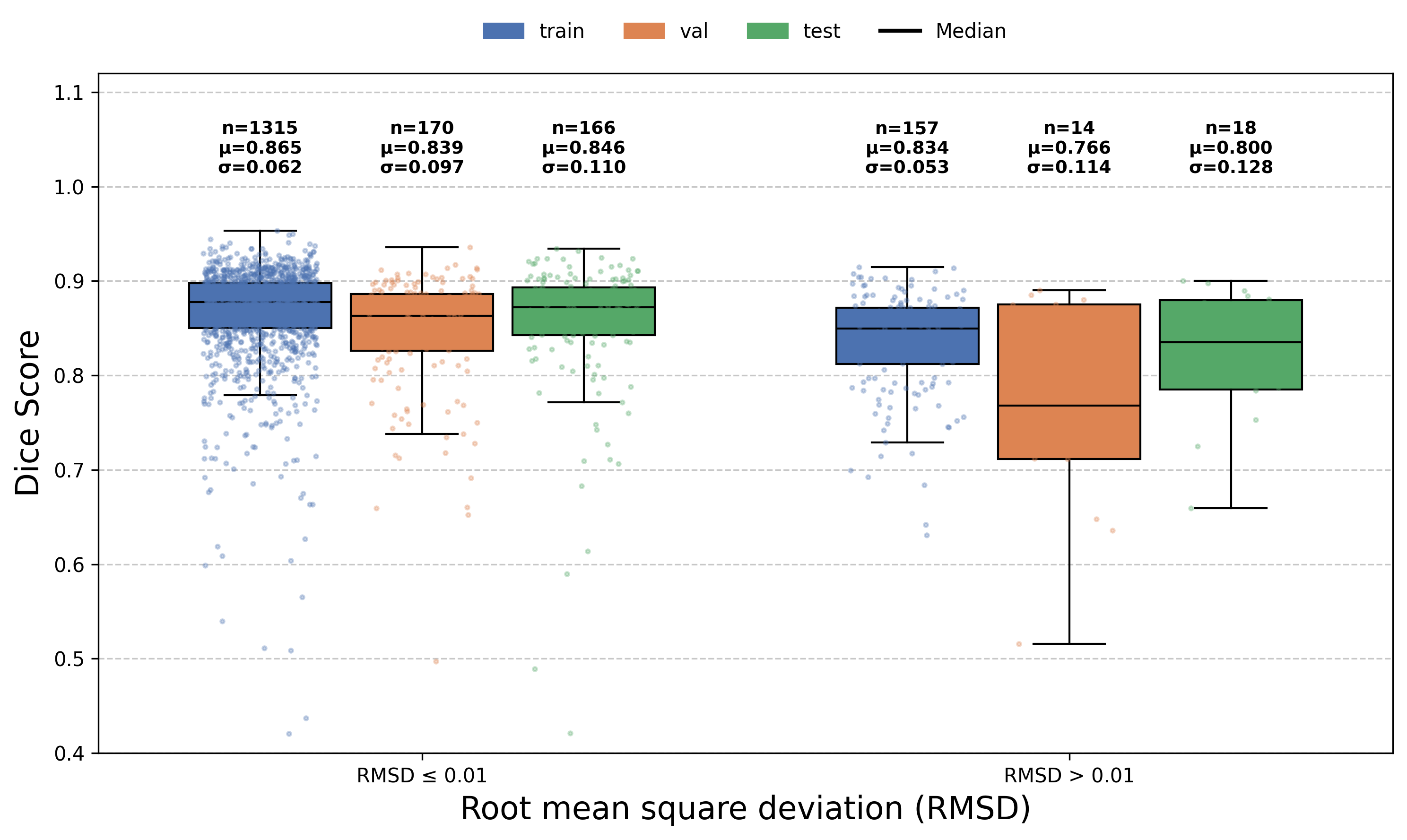}
    \caption{Model performance as a function of planarity. Molecules that deviate more from being flat ($\text{RMSD} > 0.01$ \AA) show a noticeable drop in DSC, indicating that the model has more difficulty with non-planar structures.}
    \label{fig:planar_performance_relationship}
\end{figure}

To understand the factors influencing prediction accuracy, we investigated its relationship with molecular planarity. We hypothesized that performance would degrade for minor structural deviations, as the underlying TERS data acquisition is highly sensitive to nanoscopic changes in tip-molecule orientation. This sensitivity arises because deviations from ideal planarity alter the local electromagnetic field enhancement, causing a loss of key spectroscopic information. Our results, illustrated in Fig.~\ref{fig:planar_performance_relationship}, confirm this hypothesis. The plot of DSC against root mean square deviation (RMSD) shows a marked performance degradation for molecules with RMSD $> 0.01$ \AA, characterized by a lower median DSC. This indicates that as molecules deviate from planarity, the model's predictions become less accurate and less reliable. To examine this effect further, we trained models on datasets including more non-planar molecules, as will be shown in Sec.~\ref{subsec:ablation_study}, confirming that higher RMSD correlates with lower DSC.

\begin{figure}[h!]
    \centering
    \includegraphics[width=0.9\linewidth]{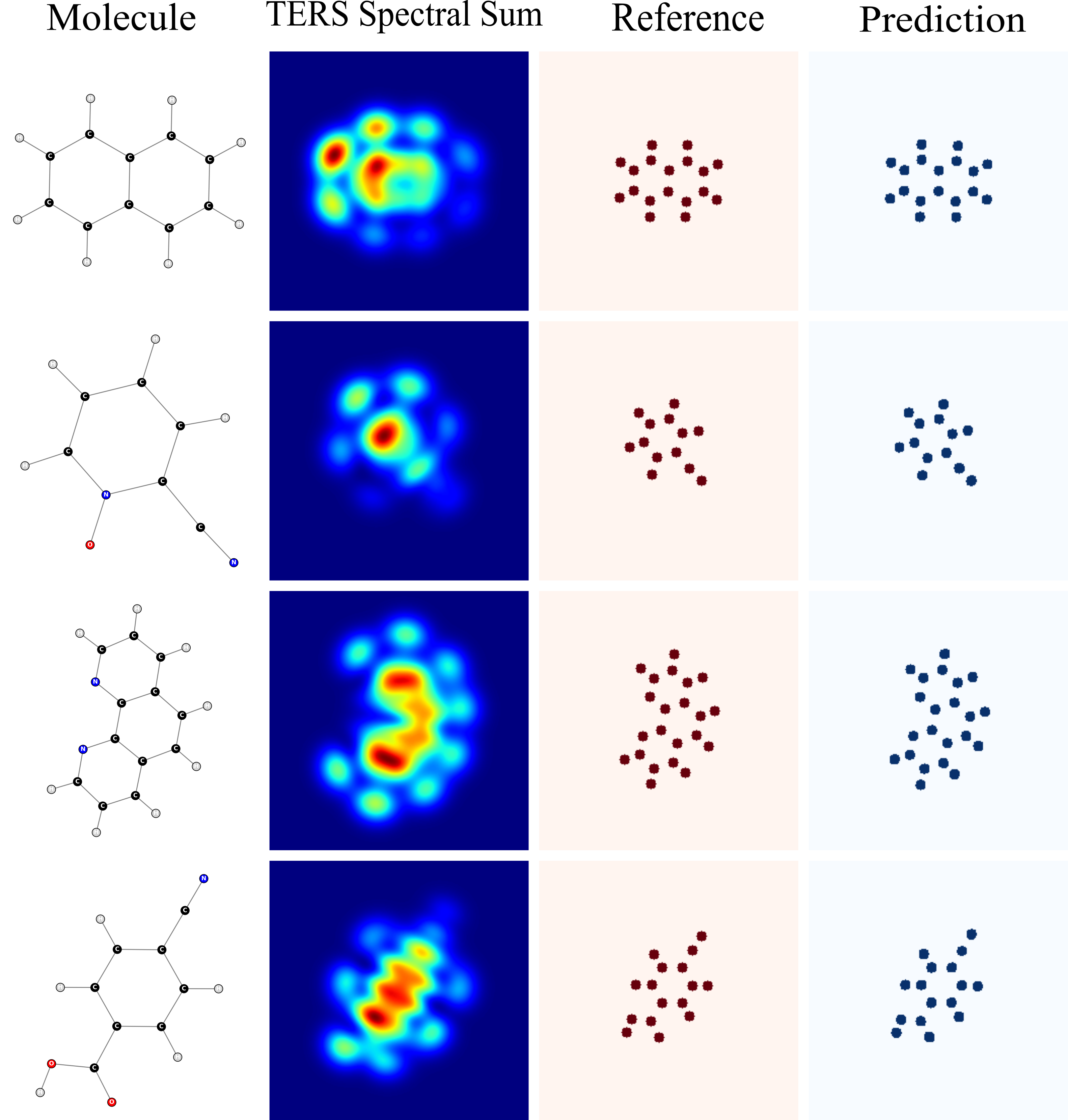}
    \caption{\textbf{Representative examples of high-fidelity predictions (top 10\% by DSC).} The model accurately reconstructs the atomic positions for molecules with strong, uniform TERS signals. Left: Ground Truth. Right: Model Prediction.}
    \label{fig:best_viz}
\end{figure}

\begin{figure}[h!]
    \centering
    \includegraphics[width=0.9\linewidth]{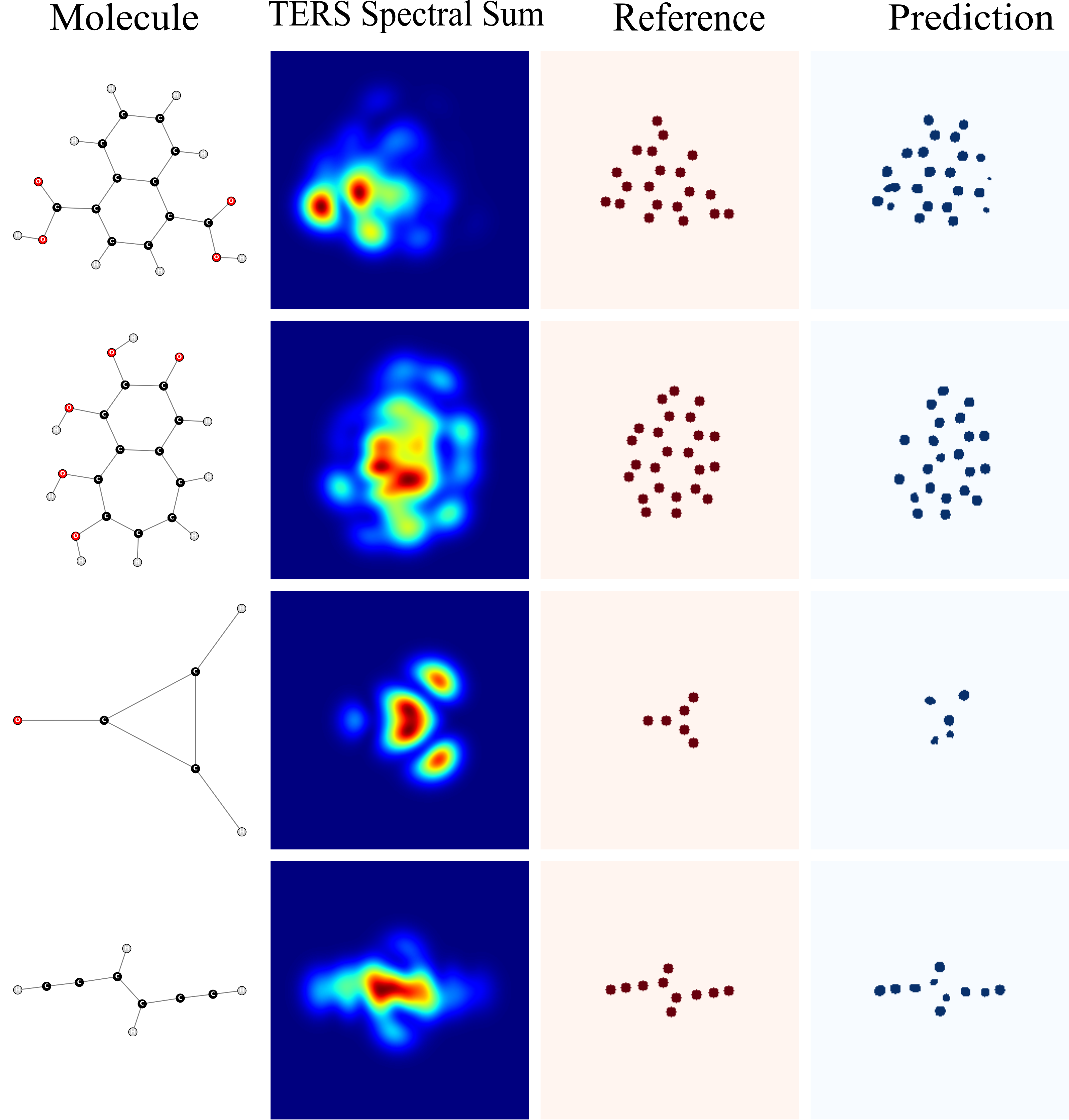}
    \caption{\textbf{Characteristic imperfect localization of the model (bottom 10\% by DSC).} Errors correlate with attenuated signals from distant atoms and geometric distortions from unequal bond signals.}
    \label{fig:worst_viz}
\end{figure}

A qualitative analysis of the top and bottom 10\% of test predictions, ordered by DSC, was performed to identify the factors influencing model performance. As shown in Figure~\ref{fig:best_viz}, predictions in the top decile exhibit high fidelity, accurately reconstructing atomic arrangements where the TERS signal is strong and uniform. Conversely, the primary causes of imperfect localization, illustrated in Figure~\ref{fig:worst_viz}, can be attributed to two main factors: (i) weak Raman signals from atoms located further from the tip, leading to reduced positional precision, and (ii) highly imbalanced Raman contribution of different bonds, which can bias the predicted geometry. Notably, even in these lower-ranked predictions, the model often generates chemically plausible structures, suggesting it has learned robust priors about molecular geometry even if the DSC is low. These limitations highlight clear directions for improving both future data acquisition protocols and model robustness.

\subsection{Ablation study on the effect of planarity on the accuracy}
\label{subsec:ablation_study}

Non-planar molecular geometries present an inherent challenge for TERS imaging because the near-field enhancement decays rapidly with increasing distance from the surface. Consequently, atomic contributions originating from out-of-plane atoms are attenuated and less consistently resolved, leading to increased variability in the measured signal for otherwise structurally similar molecules. We refer to this effect throughout as \emph{planarity-induced variability} (PIV).

\begin{figure}[h!]
    \centering

    \begin{subfigure}[b]{0.48\linewidth}
        \centering
        \includegraphics[width=\linewidth]{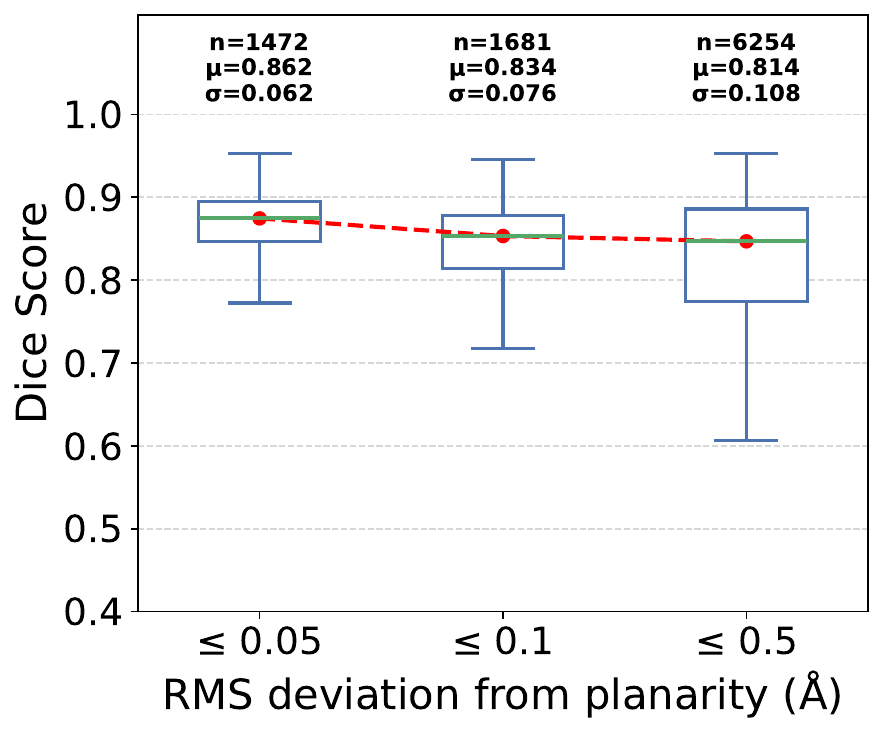}
        \caption{Training Set}
        \label{fig:dice_train}
    \end{subfigure}
    \hfill
    \begin{subfigure}[b]{0.48\linewidth}
        \centering
        \includegraphics[width=\linewidth]{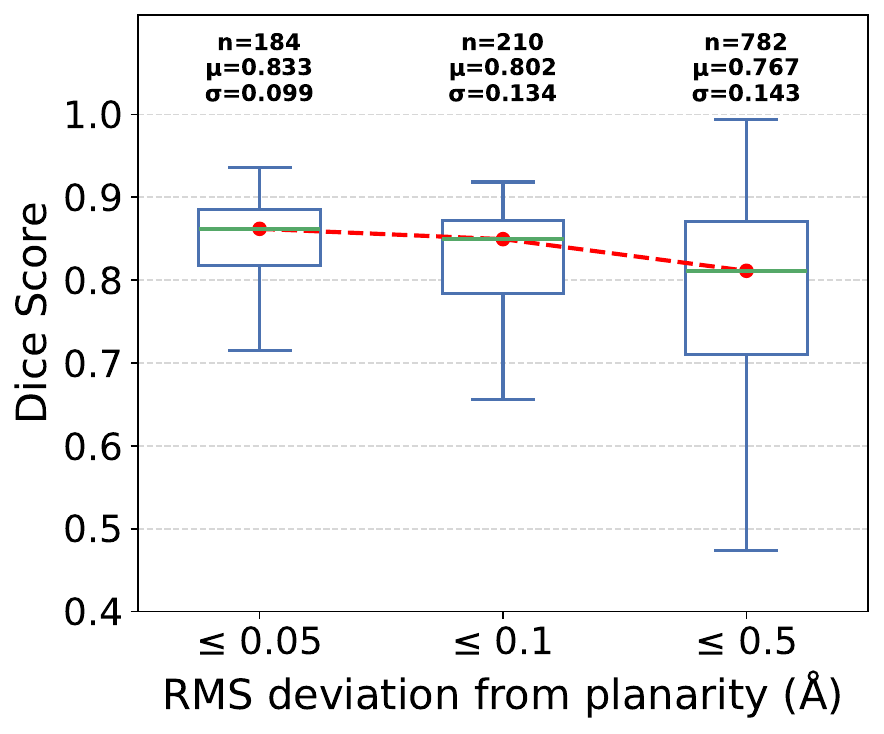}
        \caption{Validation Set}
        \label{fig:dice_val}
    \end{subfigure}

    \vspace{0.5em}
    \begin{subfigure}[b]{0.48\linewidth}
        \centering
        \includegraphics[width=\linewidth]{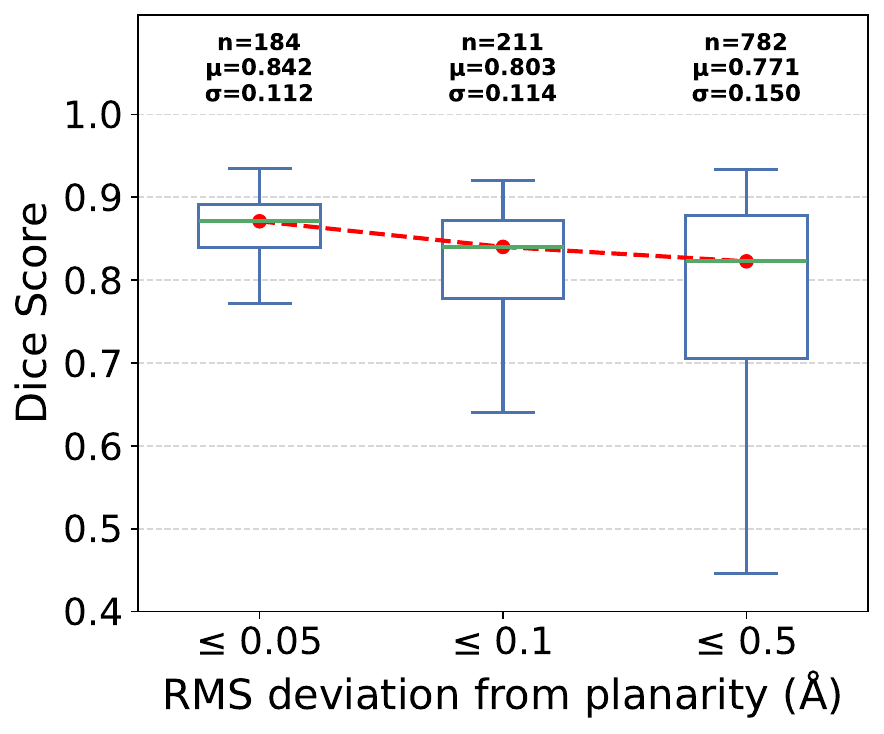}
        \caption{Test Set}
        \label{fig:dice_test}
    \end{subfigure}

    \caption{\textbf{Impact of molecular planarity on DSC.} 
    The x-axis represents a cumulative threshold ($\text{RMSD} < x$), where each bin includes all molecules with a planarity deviation less than the specified value in \AA . 
    The median trend line shows the decrease in the accuracy with increase in non-planarity.  
    As the inclusion threshold increases (allowing less planar molecules), the variance in performance increases.}
    \label{fig:dice_analysis_ext_data}
\end{figure}

To investigate the influence of PIV on model performance, datasets were constructed using progressively relaxed planarity constraints. Three models were independently trained, validated, and tested on datasets defined by RMSD thresholds of $\leq 0.05$, $\leq 0.1$ and $\leq 0.5$~\AA, enabling a controlled investigation of increasing deviations from molecular planarity. As illustrated in Fig.~\ref{fig:dice_analysis_ext_data}, model performance degrades systematically as the fraction of non-planar molecules increases: median DSC declines while the spread of the performance distribution grows. This pattern is consistent across training, validation and test splits.

Although an increase in sample size would typically be expected to reduce variance under purely statistical assumptions, the observed increase in performance spread reflects the effect of PIV on the observed TERS signal intensity. Out-of-plane distortions alter the effective signal formation by weakening or obscuring near-field contributions. As a result, the mapping from molecular structure to measured signal becomes less deterministic. In other words, PIV increases the ambiguity of the inverse problem, which manifests as higher prediction variability for models trained on datasets containing greater structural planarity variation. This analysis highlights an important practical consideration for dataset curation in TERS-based machine learning. RMSD thresholds should be chosen to balance structural diversity against the physical limitations of near-field sensitivity: including molecules with pronounced out-of-plane distortion increases PIV and degrades model performance for the present modelling approach. 

Beyond PIV, TERS presents additional constraints arising from vibrational selection effects. Stretching modes are often weak or absent, and vibrational modes oriented predominantly parallel to the surface contribute less strongly to the measured signal, whereas modes normal to the surface or aligned with the tip axis dominate the response. This introduces ambiguity in defining the ground-truth image, since atomic groups exhibiting primarily in-plane vibrations may remain weakly represented or undetected in the measurements. Consequently, certain structural regions may appear absent in the signal despite being physically present, introducing inconsistencies that can further complicate model training.

\subsection{Chemically-resolved structure prediction}
\label{sec:chem_prediction}

We further evaluate our Attention U-Net model on the joint task of atomic position and chemical species prediction for planar molecules. Whereas the atomic position prediction model (Sec.~\ref{sec:evaluation}) outputs a single foreground channel, here the network outputs one channel per element (H, C, N, O), so that each atom is represented as a circle of fixed radius assigned to its corresponding elemental channel, jointly encoding location and identity. The model achieves mean DSC values of 0.832, 0.802 and 0.810 on the training, validation and test sets respectively (Table~\ref{tab:dice_scores}), demonstrating that the architecture can resolve elemental identity in addition to molecular geometry. These values are modestly lower than the values obtained for atomic position prediction alone, consistent with the greater difficulty of the joint task, where the model must additionally discriminate between chemically distinct atoms rather than only localizing them. As in the atomic position prediction task, performance remains consistent across the three splits, indicating robust generalization.

\begin{table}[htbp]
\centering
\caption{Per-element Dice similarity coefficient (DSC) and overall mean DSC on the training, validation, and test sets.}
\label{tab:dice_scores}
\begin{tabular}{lccc}
\hline
 & Train & Validation & Test \\
\hline
H    & 0.869 & 0.858 & 0.856 \\
C    & 0.842 & 0.816 & 0.827 \\
N    & 0.782 & 0.733 & 0.759 \\
O    & 0.834 & 0.803 & 0.799 \\
\hline
Mean & 0.832 & 0.802 & 0.810 \\
\hline
\end{tabular}
\end{table}

\begin{figure}
    \centering
    \includegraphics[width=\linewidth]{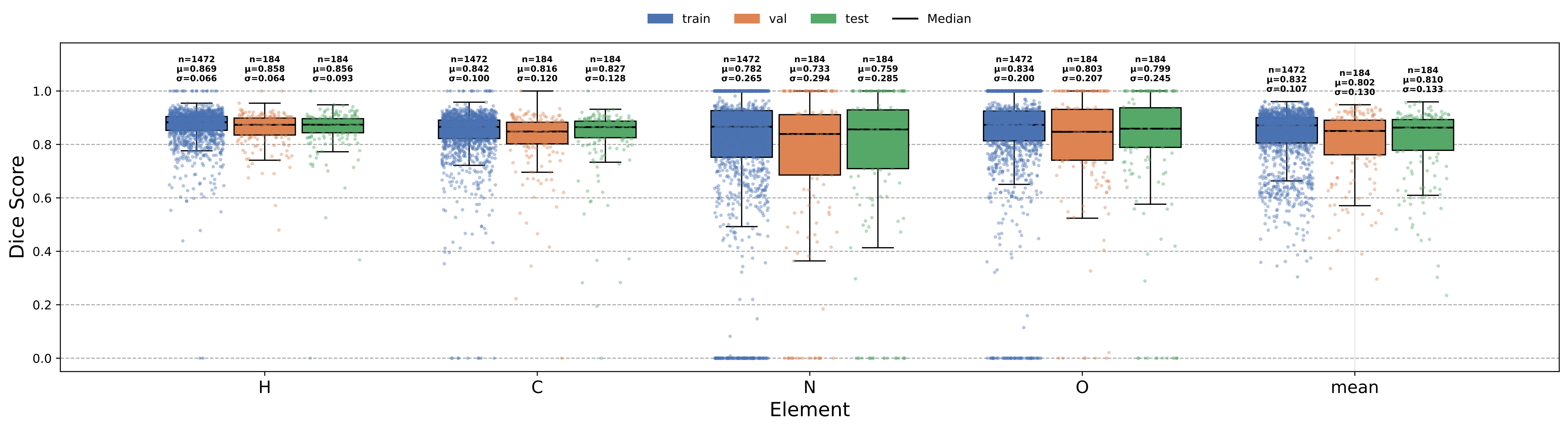}
    \caption{Distribution of per-element DSC on the test set for hydrogen, carbon, nitrogen, and oxygen. Nitrogen shows both the lowest median DSC and the largest spread, while hydrogen and carbon are predicted most consistently. This trend mirrors the relative abundance of each element in the molecule dataset (Fig.~\ref{fig:mol_dataset}).}
    \label{fig:chem_dice}
\end{figure}

The dataset is strongly imbalanced in chemical composition, with hydrogen and carbon far more abundant than nitrogen and oxygen. For evaluation, we report the mean of the per-element DSC values (Table~\ref{tab:dice_scores}). As shown in Fig.~\ref{fig:chem_dice}, hydrogen and carbon are resolved most accurately, whereas nitrogen attains both the lowest median DSC and the highest variance, giving a spread of roughly 0.10 in DSC between the best and worst element on the test set. This per-element trend in both mean and variance correlates directly with the relative abundance of each element in the dataset (Fig.~\ref{fig:mol_dataset}).

\begin{figure}
    \centering
    \includegraphics[width=0.9\linewidth]{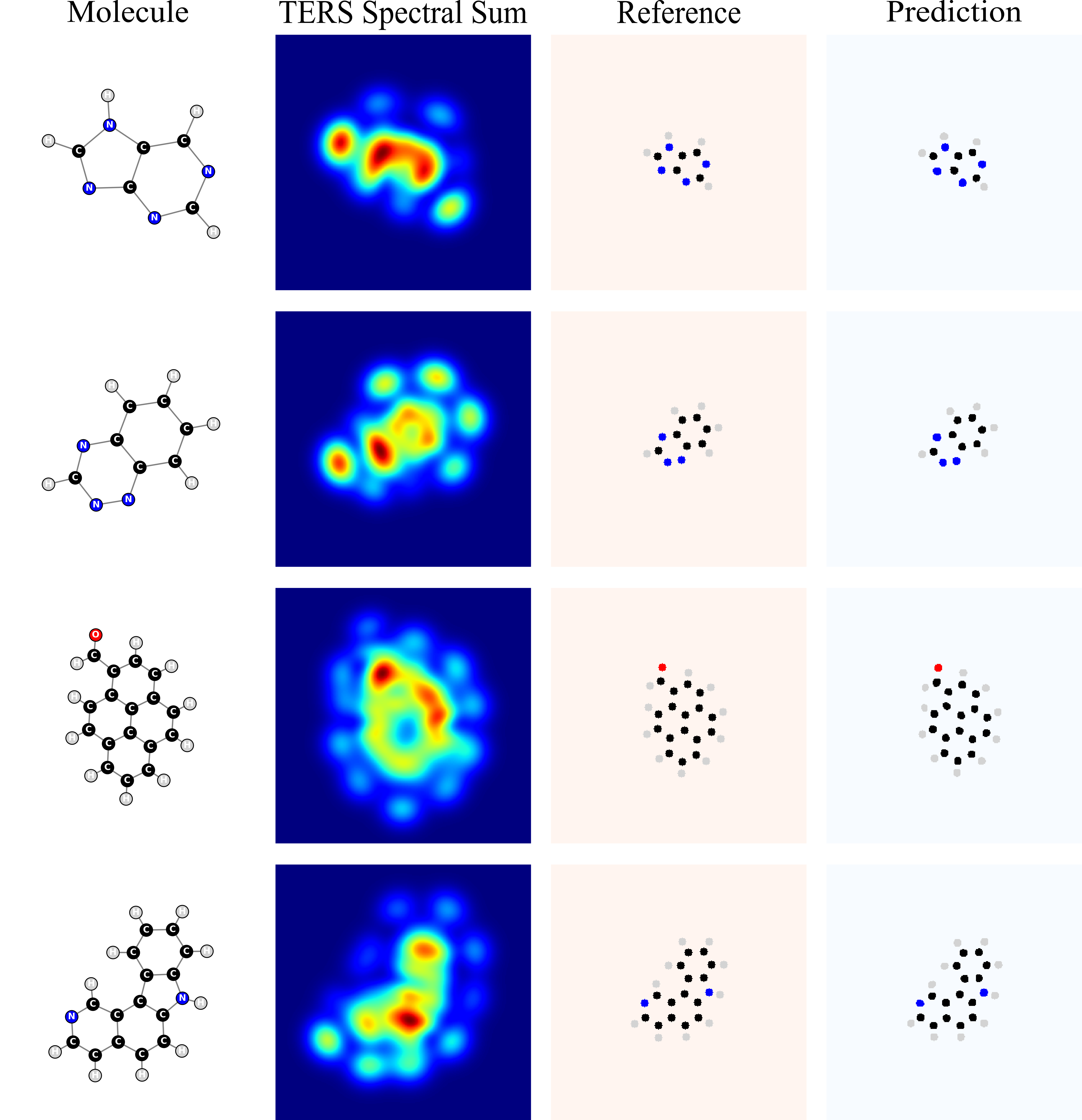}
    \caption{Representative examples of high-fidelity chemical species predictions (top 10\% by DSC). The model accurately reconstructs both the atomic positions and elemental identities (H, C, N, O) for molecules with strong, uniform TERS signals. Left: Ground Truth. Right: Model Prediction.}
    \label{fig:chem_best_viz}
\end{figure}

\begin{figure}
    \centering
    \includegraphics[width=0.9\linewidth]{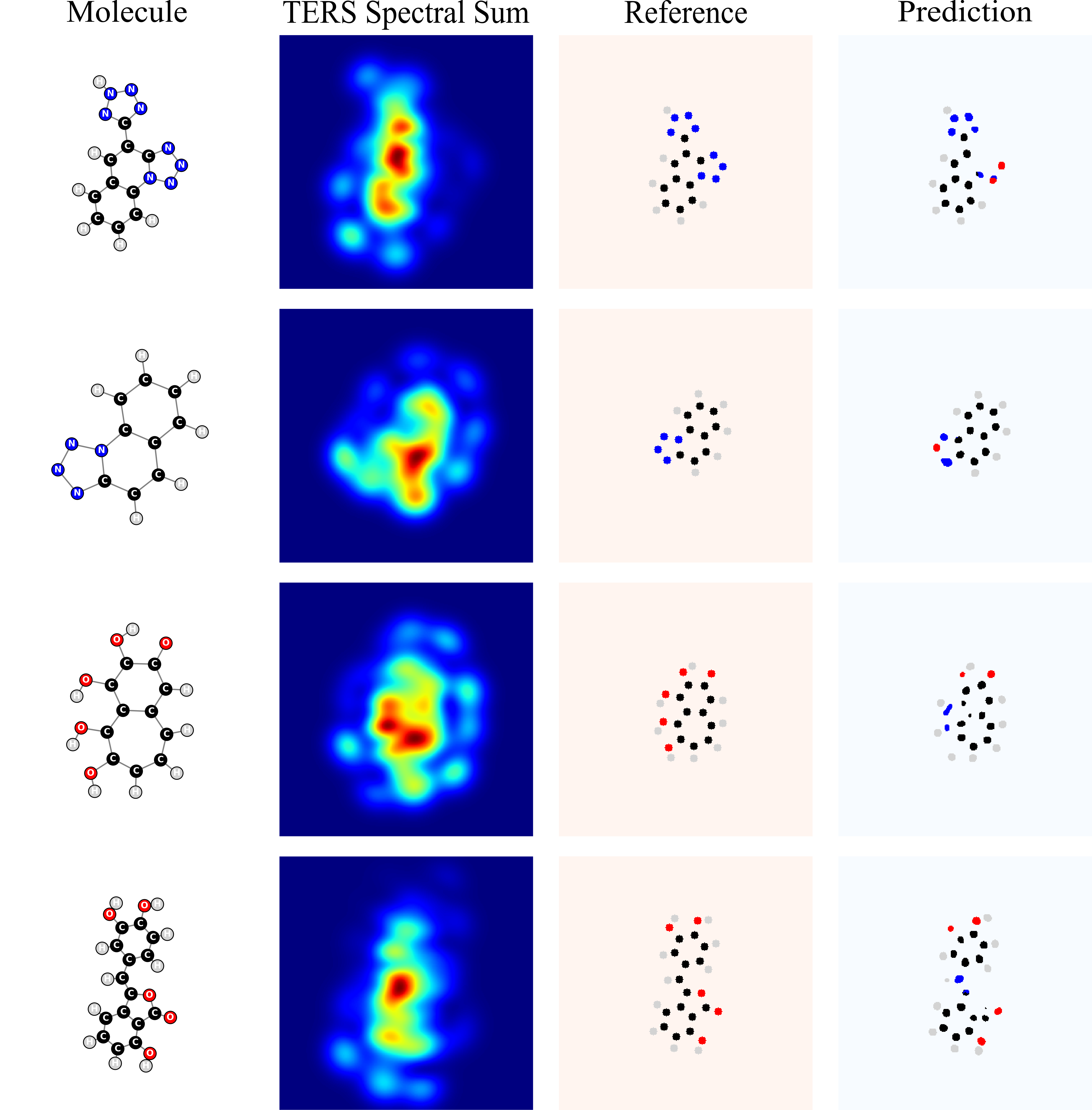}
    \caption{Representative examples of low-fidelity chemical species predictions (bottom 10\% by DSC). Carbon and hydrogen atoms remain well localized, but predictions for oxygen and nitrogen are frequently misplaced or misassigned, reflecting their underrepresentation in the training dataset. Left: Ground Truth. Right: Model Prediction.}
    \label{fig:chem_worst_viz}
\end{figure}

To further characterize model behavior, we visualize the top and bottom 10\% of test-set predictions, ranked by overall DSC. As shown in Fig.~\ref{fig:chem_best_viz}, predictions in the top decile closely reproduce the ground-truth atomic maps for all four elements, confirming the model's ability to jointly resolve molecular geometry and chemical identity under favorable conditions. The bottom decile (Fig.~\ref{fig:chem_worst_viz}), in contrast, highlights the main limitations of the current model. A closer inspection of these cases shows that, even when the overall TERS signal is strong, predictions for carbon and hydrogen remain accurate, whereas predictions for oxygen and nitrogen are markedly less reliable. This reinforces the need to enrich the training set with molecules containing a higher proportion of nitrogen and oxygen, exposing the model to a broader range of functional groups and thereby improving the accuracy of chemical species identification.

\section{Conclusions}
\label{sec:conclusion}

In this work, we introduced a deep learning framework termed SMARTERS capable of predicting molecular structures from simulated TERS hyperspectral images. By training an encoder-decoder network to map vibrational spectra to atomic coordinates, we demonstrate high fidelity in reconstructing planar molecular geometries from simulated images, going beyond simpler classification challenges \cite{tian_autonomous_2026}. This result establishes the viability of a ML-driven approach to bypass traditional, often ambiguous, manual interpretation, thereby opening a new avenue for automated molecular discovery from TERS.

While this study serves as a critical proof of concept, we identify three key areas for future research. First, extending the approach to non-planar systems remains an open challenge. The core difficulty is that atoms located beneath the surface do not contribute directly to the TERS signal but still influence the vibrational modes through their bonds. We posit that a transition from 2D representations to 3D molecular graphs, processed by graph neural networks, presents the most compelling path forward. Such a framework is not only essential for inferring the complete 3D geometry from the indirect vibrational contributions of subsurface atoms, but would also inherently provide robustness to molecular orientation and help disentangle the true signal from tip-positional artifacts.
Second, while we have demonstrated chemically-resolved prediction for H, C, N, and O, its reliability on the rarer species (N, O) is currently limited by the chemical imbalance of our dataset. Extending this capability will require curating datasets with greater elemental and functional-group diversity, moving toward complete, chemically-resolved structure determination.
Finally, bridging the sim-to-real gap is paramount for practical application and the current model clearly is not yet capable of routinely handling experimental data (see App. \ref{sec:experimental}). Experience from ML models in SPM suggest that the model's ability to generalize to diverse experimental TERS data is contingent on training with larger datasets, enriched with complex augmentations that realistically simulate a wide variety of experimental conditions, including thermal noise, tip variations, and substrate effects (see also \cite{cui_ters-abnet_2026}).

\section{Acknowledgements}

The authors thank Krystof Brezina and Mariana Rossi for insightful discussions and helpful feedback. The authors also thank prof. Fernando Aguilar-Galindo, prof. Martin \v{S}vec, and prof. Pablo Merino for providing original experimental data. The work was supported by the Finnish Ministry of Education and Culture through the Quantum Doctoral Education Pilot Program (QDOC VN/3137/2024-OKM-4) and the Research Council of Finland through the Finnish Quantum Flagship project (358877, Aalto University). O.J.S. acknowledge the financial support from the Research Council of Finland No. 371875. The authors acknowledge the computational resources provided by the Aalto Science-IT Project and CSC. 

\section{Data availability}

The data that support the findings of this study are openly available in Zenodo \cite{zenodo_smarters}. The scripts used for data analysis are publicly available on GitHub \cite{github_smarters}. 

\input{X_suppl}



\end{document}

%% file: X_suppl.tex
\clearpage
\appendix

\section*{Appendices}
\addcontentsline{toc}{section}{Appendix}


\section{Model}
\label{sec:model}

\begin{figure}[h]
    \centering
    \includegraphics[width=0.8\linewidth]{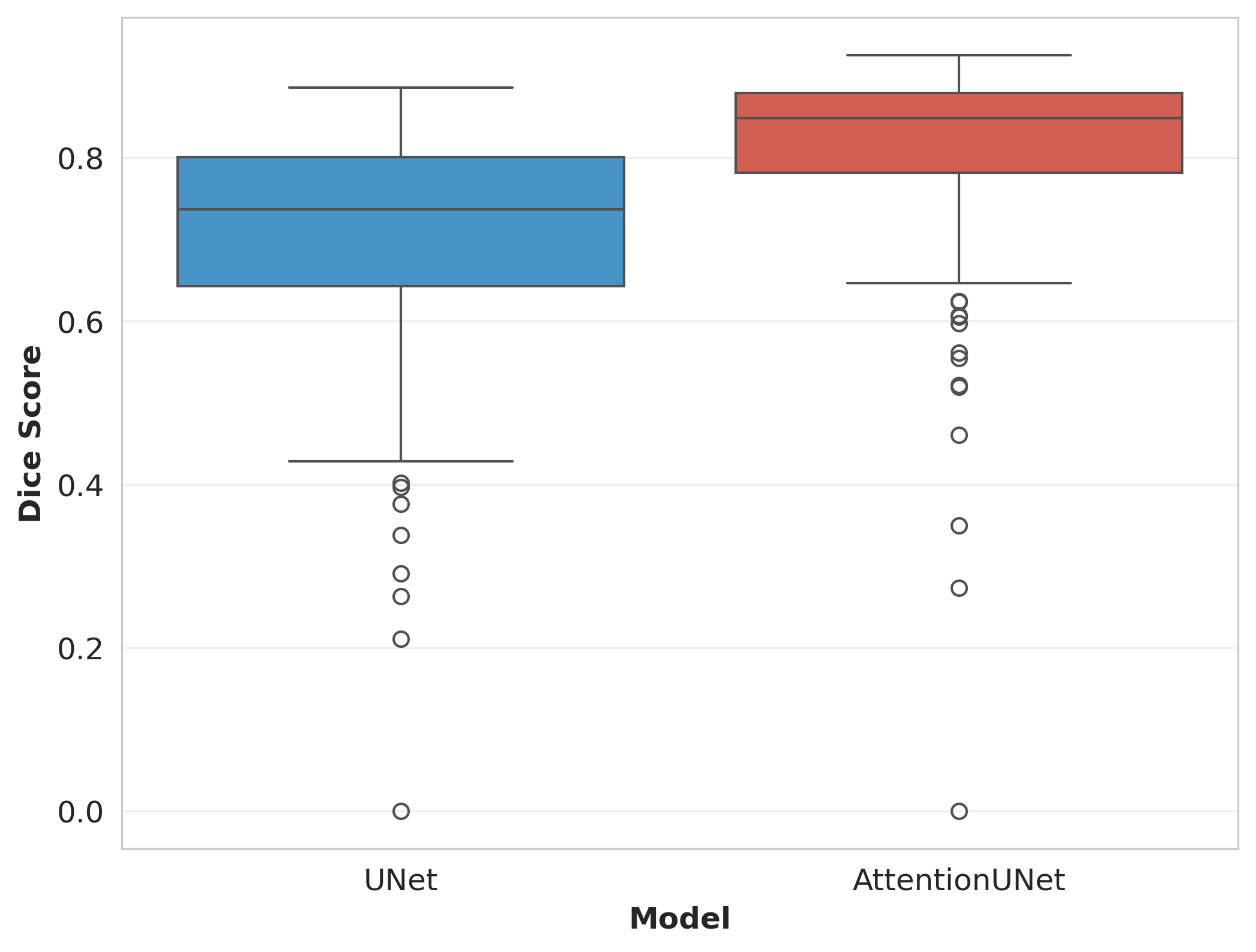}
    \caption{Performance comparison between standard U-Net and Attention U-Net architectures, on TERS data. }
    \label{fig:model_comparison}
\end{figure}

\begin{figure}
    \centering
    \includegraphics[width=0.99\linewidth]{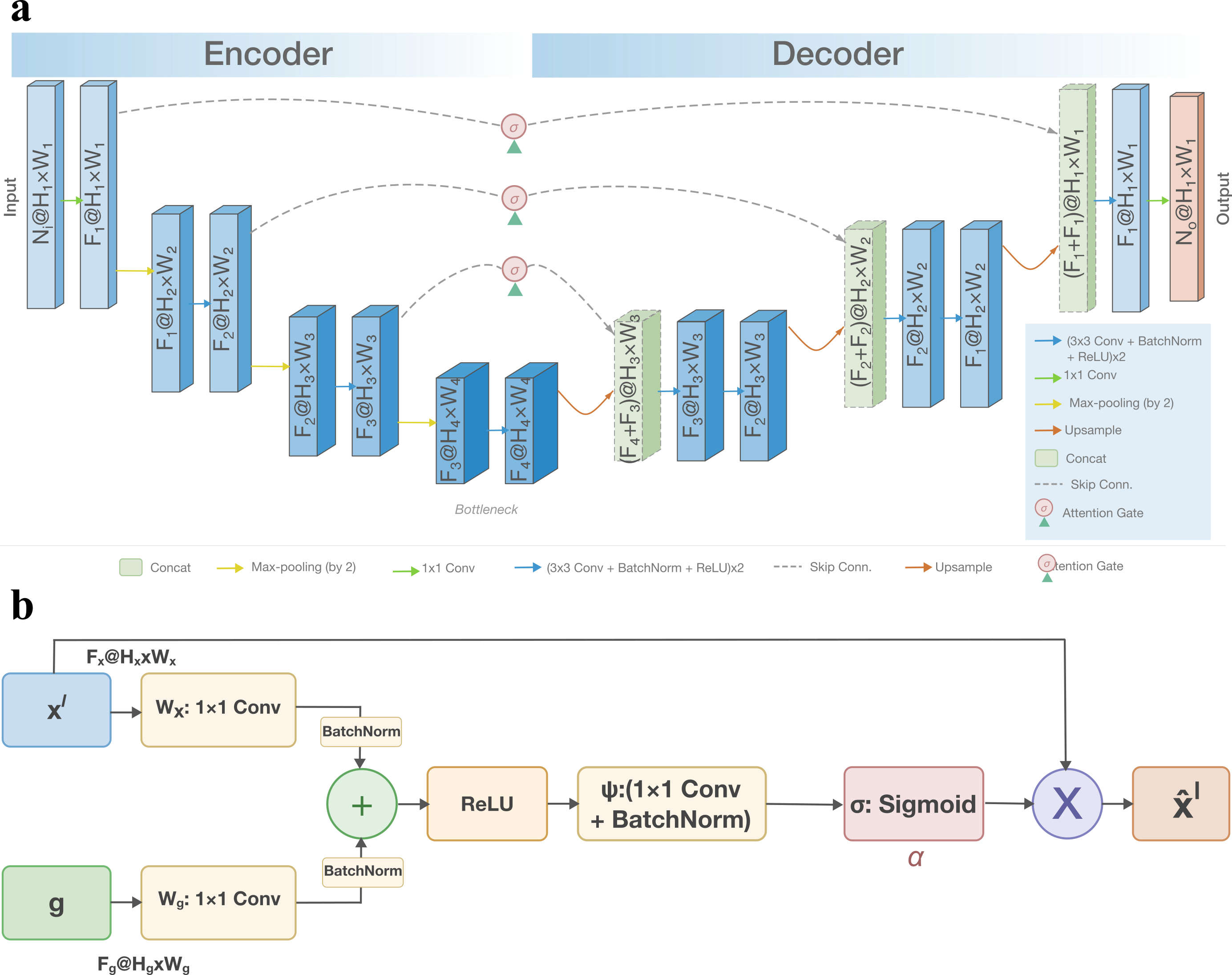}
    \caption{
(a) Attention U-Net architecture used for structure prediction from TERS images, illustrated here with a 3-level encoder/decoder depth, together with the additive attention gate applied on the skip connections.
$F_k$ denotes the number of filters at stage $k$, while $N_i$ and $N_o$ denote the input and output channels, respectively; $N_o$~= {1,4}  depending on the task (structure vs. chemical species prediction) 
(b) Additive attention gate for skip-connection filtering.
Decoder context $g$ is used to compute attention coefficients $\alpha_l$ that scale encoder features $x_l$, suppressing irrelevant activations prior to concatenation in the decoder.
}

    \label{fig:attention_unet_full}
\end{figure}

The encoder-decoder structure used for structure discovery for TERS is Attention U-Net model originally introduced by Oktay \textit{et al.} \cite{oktay2018attention}. It uses U-Net architecture \cite{ronneberger2015u} with attention mechanism \cite{vaswani2023attentionneed} and achieves higher accuracy than the U-Net baseline, as shown in Fig.~\ref{fig:model_comparison}. An overview of the network and the additive attention gate module is shown in Fig.~\ref{fig:attention_unet_full}. The model operates on 2-D TERS images and follows a symmetric encoder--decoder design (Fig.~\ref{fig:attention_unet_full}a). Each encoder stage consists of two successive $3\times3$ convolutional layers with batch normalisation and ReLU activation, followed by max pooling with a downsampling factor of $2$.

In the encoder, the model learns hierarchical features by progressively reducing the spatial resolution while increasing the number of feature channels. The decoder mirrors this process using bilinear upsampling, concatenation with the corresponding encoder features, and convolutional refinement to recover spatial detail (Fig.~\ref{fig:attention_unet_full}a). The encoder--decoder depth and channel widths used in each configuration are reported in Table~\ref{tab:hyperparams} for the atomic structure prediction model and in Table~\ref{tab:optuna_none_v2_search} for the chemically-resolved prediction model. To suppress irrelevant encoder activations, skip connections are filtered using additive attention gates (Fig.~\ref{fig:attention_unet_full}b). Let $x_l \in \mathbb{R}^{C_l \times H_l \times W_l}$ denote the encoder feature map at level $l$, and let $g \in \mathbb{R}^{C_g \times H_l \times W_l}$ denote the decoder ``gating'' feature map from a coarser scale, resized to match the spatial resolution $(H_l, W_l)$. Both are linearly projected using $1\times1$ convolutions into a common intermediate space of $C_{\mathrm{att}}$ channels and combined to compute attention coefficients
\begin{equation}
\alpha_l = \sigma\!\left(\psi^\top \,\mathrm{ReLU}\!\left(W_x x_l + W_g g + b\right)\right),
\end{equation}

\noindent where $W_x$ and $W_g$ are learned $1\times1$ convolutional projections, $b$ is a bias term, $\mathrm{ReLU}(\cdot)$ is the rectified linear unit, and $\psi^\top$ denotes a learned $1\times1$ projection mapping the intermediate representation to a single-channel attention logit. The sigmoid activation produces an attention map
$\alpha_l \in [0,1]^{1 \times H_l \times W_l}$.
This map is applied channel-wise to the encoder features,
yielding the gated representation
$\tilde{x}_l = \alpha_l \odot x_l$,
where $\odot$ denotes element-wise multiplication. The gated features $\tilde{x}_l$ are concatenated with decoder features before convolutional refinement. This mechanism allows decoder context to emphasise spatial regions consistent with the predicted molecular structure while suppressing background or noisy artefacts in simulated TERS images.

In contrast to the original medical-imaging implementation, our model processes 2-D inputs with configuration-dependent depth and channel widths. A final $1\times1$ convolution maps the last decoder features to the output map, whose number of channels $N_o$ depends on the task: $N_o=1$ for the atomic structure prediction model (a single structure-vs-background mask) and $N_o=4$ for the chemically-resolved prediction model (one channel per element: H, C, N, O).
The resulting architecture retains U-Net’s localisation capability, while attention gating conditions skip-feature transfer on decoder context, enabling selective integration of spatial detail relevant to the predicted structure.

\begin{table}[htbp]
  \centering
  \captionsetup{width=.9\textwidth, justification=justified}
  \caption{Hyperparameter search space and selected model configurations for the \emph{atomic structure prediction} model, for datasets constructed using different RMSD thresholds. For each threshold, we report the configuration achieving the best validation performance for the model trained on the corresponding dataset. Filter configurations define both the number of filters per stage and the encoder/decoder depth: \textsf{A}~=~$[16,32,64]$~(3-level), \textsf{B}~=~$[16,32,64,128]$~(4-level), \textsf{C}~=~$[16,32,64,128,256]$~(5-level), \textsf{D}~=~$[32,64,128,256,512]$~(5-level).}
  \label{tab:hyperparams}
  \small
  \renewcommand{\arraystretch}{1.25}
  \setlength{\tabcolsep}{8pt}        
  \begin{tabular}{@{} l @{\hspace{12pt}} l @{\hspace{14pt}} c @{\hspace{14pt}} c @{\hspace{14pt}} c @{}}
    \toprule
                                  &
                                  & \multicolumn{3}{c}{\textbf{Selected value}} \\
                                    \cmidrule(l){3-5}
    \textbf{Hyperparameter}       & \textbf{Search space}
                                  & \textbf{0.05\,\AA}
                                  & \textbf{0.10\,\AA}
                                  & \textbf{0.50\,\AA}                               \\
    \midrule
    Input channels                & $\{100,\; 400\}$
                                  & $100$  & $100$  & $400$                          \\
    Encoder/decoder filters       & $\{\textsf{A},\;\textsf{B},\;\textsf{C},\;\textsf{D}\}$
                                  & \textsf{C}  & \textsf{B}  & \textsf{D}          \\
    Attention channels            & $\{16,\; 32,\; 64\}$
                                  & $16$   & $16$   & $32$                           \\
    Batch size                    & $\{16,\; 32,\; 64,\; 128\}$
                                  & $16$   & $32$   & $16$                           \\
    Learning rate                 & $[10^{-5},\; 10^{-3}]$
                                  & $7.35\!\times\!10^{-4}$
                                  & $5.56\!\times\!10^{-4}$
                                  & $1.29\!\times\!10^{-4}$                          \\
    \bottomrule
  \end{tabular}
\end{table}

\begin{table}[htbp]
  \centering
  \captionsetup{width=.9\textwidth, justification=justified}
  \caption{Hyperparameter search space and selected configuration for the \emph{chemically-resolved prediction} model, trained on the 0.05~\AA{} dataset. Only the encoder/decoder filters and the learning rate were searched; the remaining hyperparameters were fixed (100 input channels, 64 attention channels, batch size 16, 50 epochs). Filter configurations define both the number of filters per stage and the encoder/decoder depth: 
  \textsf{A}~$=$~$[32,64,128,256]$, 
  \textsf{B}~$=$~$[64,128,256]$, 
  \textsf{C}~$=$~$[64,128,256,512]$, 
  \textsf{D}~$=$~$[32,64,128,256,512]$, 
  \textsf{E}~$=$~$[64,128,256,512,1024]$, 
  \textsf{F}~$=$~$[64,128,256,256]$.}
  \label{tab:optuna_none_v2_search}
  \small
  \renewcommand{\arraystretch}{1.25}
  \setlength{\tabcolsep}{8pt}
  \begin{tabular}{@{} l l c @{}}
    \toprule
    \textbf{Hyperparameter} & \textbf{Search space} & \textbf{Best value} \\
    \midrule
    Encoder/decoder filters &
    \{\textsf{A}, \textsf{B}, \textsf{C}, \textsf{D}, \textsf{E}, \textsf{F}\} &
    \textsf{D} \\
    Learning rate &
    $[10^{-5},\,10^{-3}]$ &
    $2.33\times10^{-4}$ \\
    \bottomrule
  \end{tabular}

\end{table}

\section{Dataset Preprocessing}
\label{sec:preprocessing}
\subsection{Principal Component Analysis for Planarity Selection and Tip Alignment}

To ensure reliable TERS simulations in the current model, only planar molecules were included in the dataset. Planarity was determined using principal component analysis (PCA) on the 3D atomic coordinates $\mathbf{X} \in \mathbb{R}^{N \times 3}$, where $N$ is the number of atoms. PCA decomposes the molecular coordinates into orthogonal axes of maximum variance by diagonalizing the covariance matrix:

\begin{equation}
\mathbf{C} = \frac{1}{N} (\mathbf{X} - \bar{\mathbf{X}})^\top (\mathbf{X} - \bar{\mathbf{X}})
\end{equation}

\noindent where $\bar{\mathbf{X}}$ is the mean atomic coordinate vector. The eigendecomposition of $\mathbf{C}$ yields three eigenvalues $\lambda_1 \geq \lambda_2 \geq \lambda_3 \geq 0$ and a corresponding orthonormal set of eigenvectors $\{\mathbf{v}_1, \mathbf{v}_2, \mathbf{v}_3\}$. These eigenvalues quantify the variance of the atomic coordinates along the principal axes defined by the eigenvectors. The plane spanned by $\{\mathbf{v}_1, \mathbf{v}_2\}$ constitutes the best-fit plane through the molecule, while $\mathbf{v}_3$ is the normal vector to this plane.

Planar molecules were identified by checking the variance along the axis perpendicular to the molecular plane. Specifically, a molecule was classified as planar if:

\begin{equation}
\lambda_3 < \epsilon
\end{equation}

\noindent where $\lambda_3$ is the smallest eigenvalue corresponding to the axis with minimal variance, and $\epsilon$ is a small threshold indicating negligible out-of-plane deviation. This ensures that most atoms lie approximately within a single plane. Equivalently, $\sqrt{\lambda_3}$ corresponds to the root-mean-square out-of-plane deviation of the atoms from the best-fit plane, and the above criterion enforces negligible displacement normal to the plane.

\begin{figure}[htb]
    \centering
    \begin{subfigure}[b]{0.32\linewidth}
        \centering
        \includegraphics[width=\linewidth]{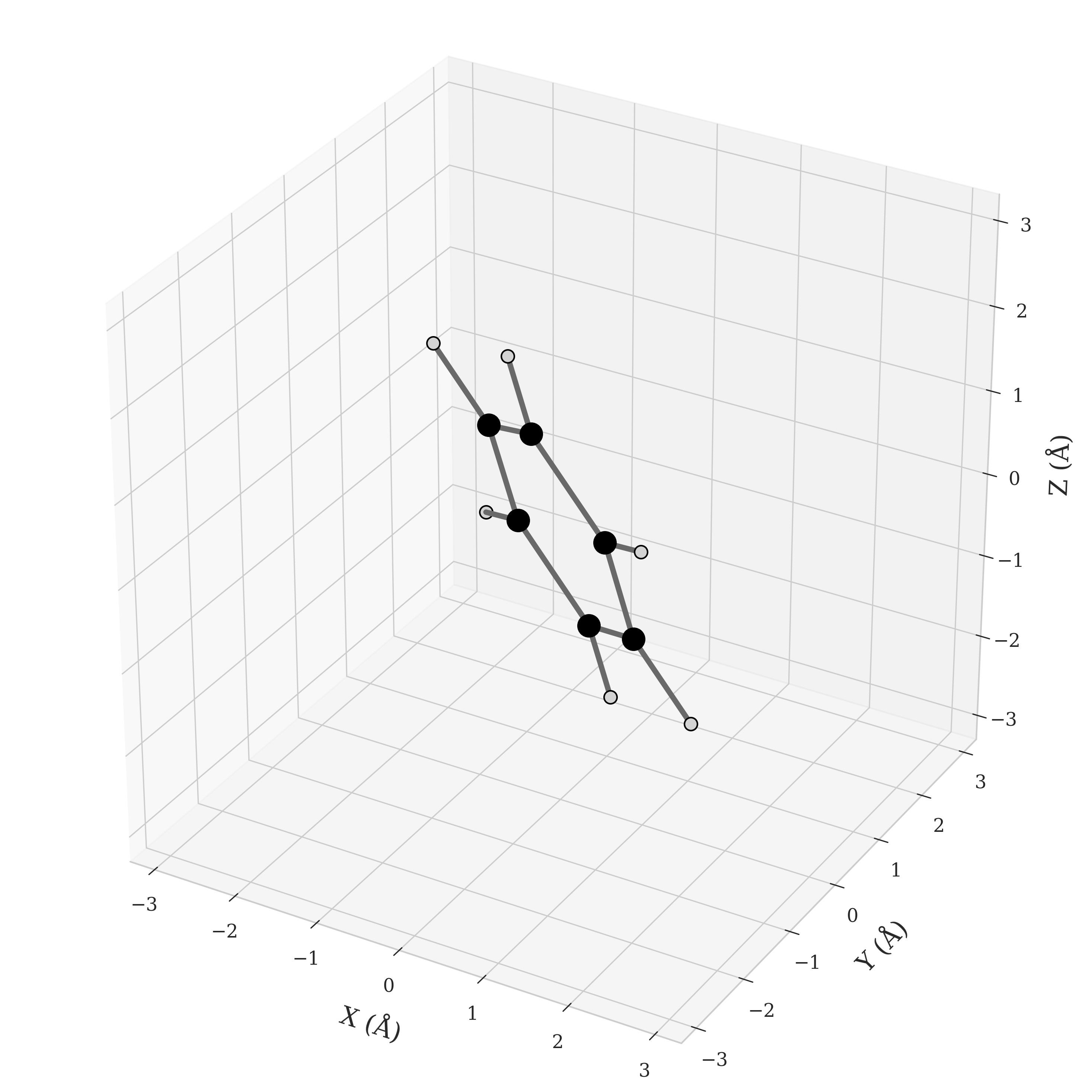}
        \caption{Original molecule}
        \label{fig:molecule_original}
    \end{subfigure}
    \hfill
    \begin{subfigure}[b]{0.32\linewidth}
        \centering
        \includegraphics[width=\linewidth]{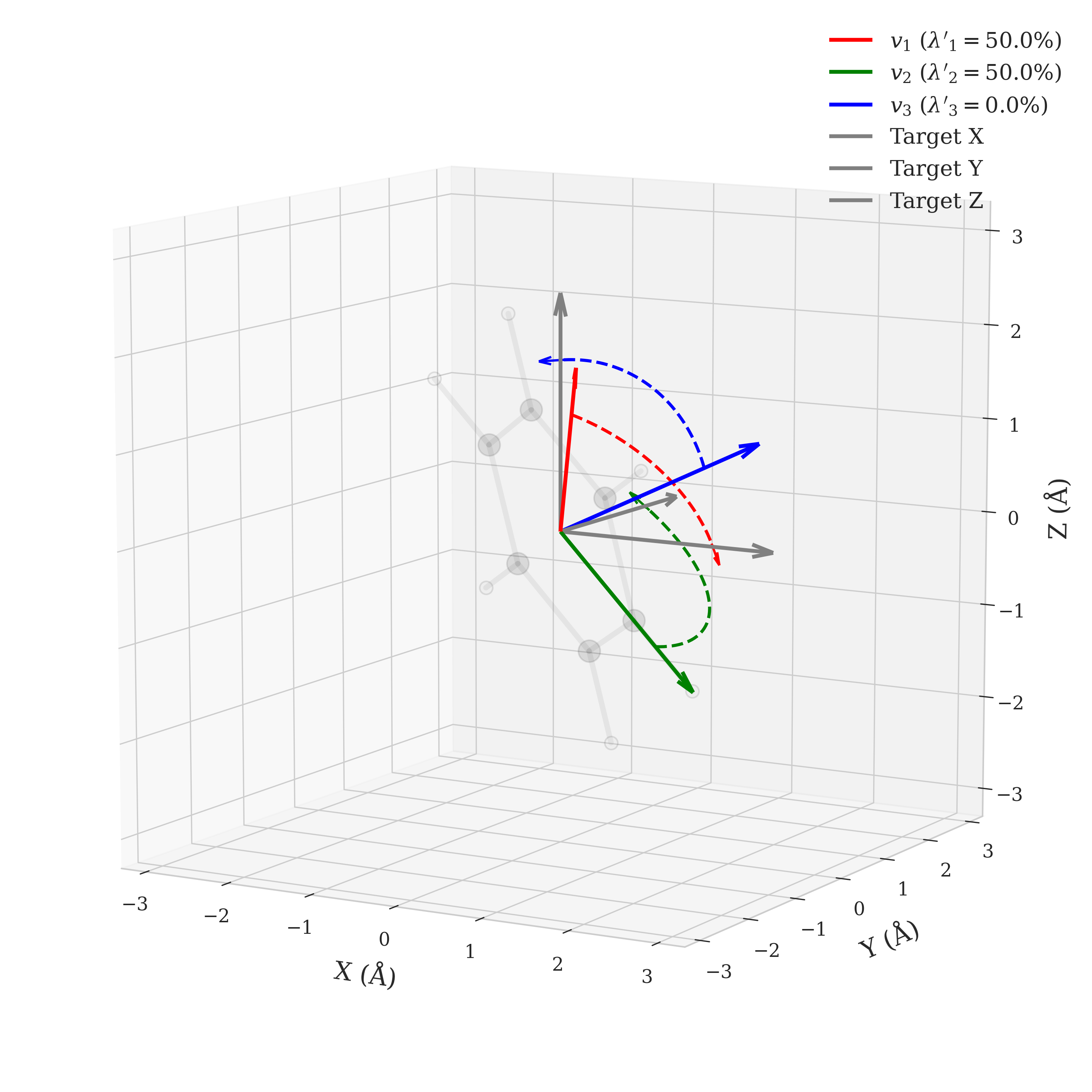}
        \caption{PCA alignment}
        \label{fig:pca_axes}
    \end{subfigure}
    \hfill
    \begin{subfigure}[b]{0.32\linewidth}
        \centering
        \includegraphics[width=\linewidth]{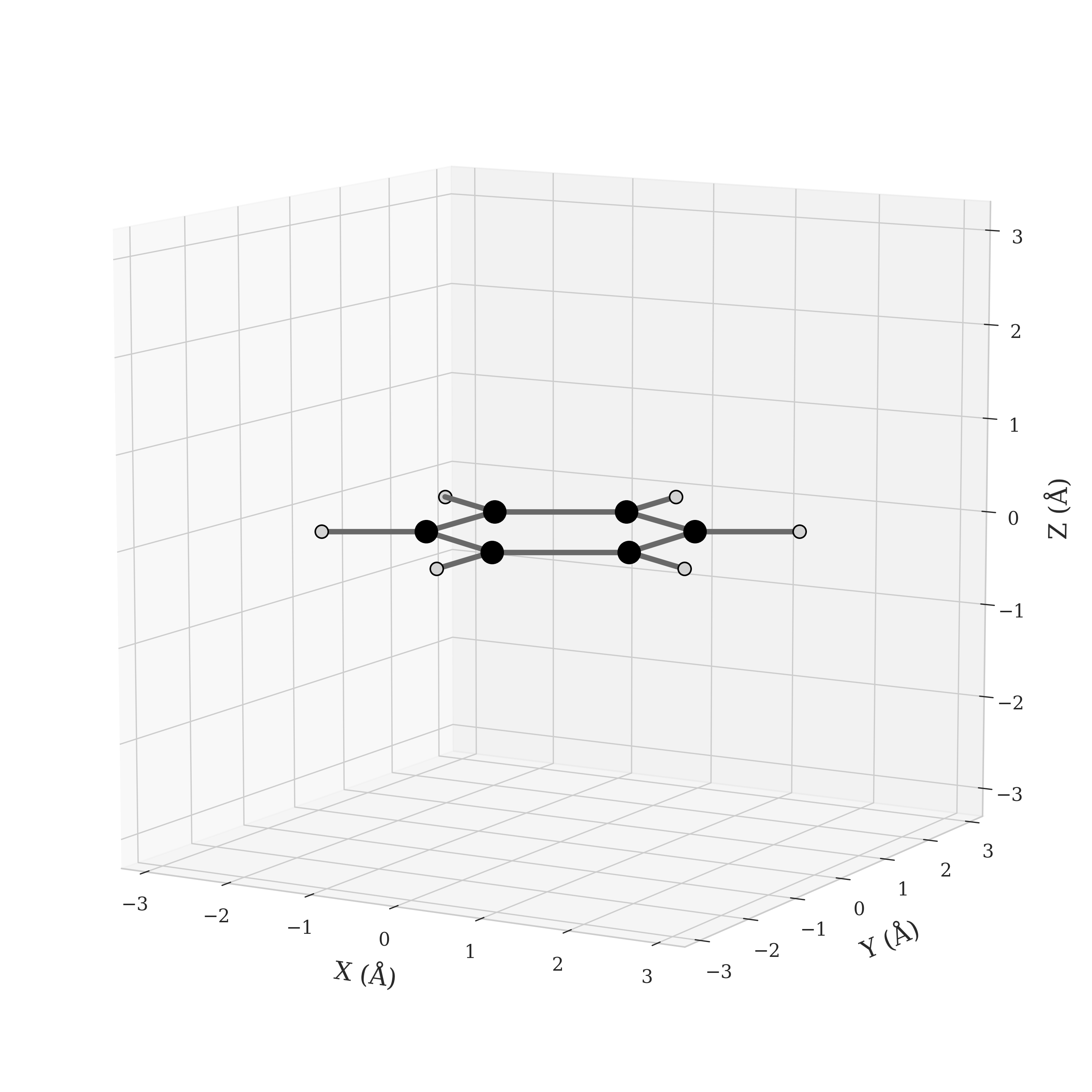}
        \caption{Rotated molecule}
        \label{fig:molecule_rotated}
    \end{subfigure}

    \caption{Dataset preprocessing pipeline. (a) Original 3D coordinates of a molecule. (b) PCA identifies principal axes and eigenvalues; molecules with minimal out-of-plane variance (\(\lambda_3 < \epsilon\)) are selected as planar. (c) Selected planar molecules are rotated using rotation matrix \(\mathbf{R}\) so the molecular plane lies in the XY-plane and the tip is along the Z-axis, ensuring consistent orientation for TERS simulations.}
    \label{fig:pca_rotation_pipeline}
\end{figure}

After planar molecules were selected, each molecule was rotated to align its plane perpendicular to the TERS tip. Let $\mathbf{R}$ be the rotation matrix constructed from PCA eigenvectors. The coordinates were transformed as:

\begin{equation}
\mathbf{X}_{\text{aligned}} = \mathbf{R} (\mathbf{X} - \bar{\mathbf{X}})
\end{equation}

\noindent The rotation aligns the molecular plane with the XY-plane and the tip along the Z-axis, providing consistent orientation across all molecules and maximal structural visibility in TERS simulations.

\section{Evaluation metrics}
\label{sec:evaluation_metrics}

\subsection{Dice Similarity Coefficient}
\label{sec:dice_score}

The Dice similarity coefficient (DSC) is closely related to Intersection over Union (IoU) but gives more weight to the overlap between prediction and ground truth. It is defined as:

\begin{equation}
\label{eq:dice}
\text{Dice}(P,G) = \frac{2|P \cap G|}{|P| + |G|}
\end{equation}

\noindent Like IoU, its value lies between 0 and 1, with higher scores indicating greater similarity. Both IoU and DSC provide complementary perspectives on segmentation accuracy, but the DSC emphasizes overlap accurately and is less sensitive to class imbalance. 

\begin{figure}
    \centering
    \includegraphics[width=\linewidth]{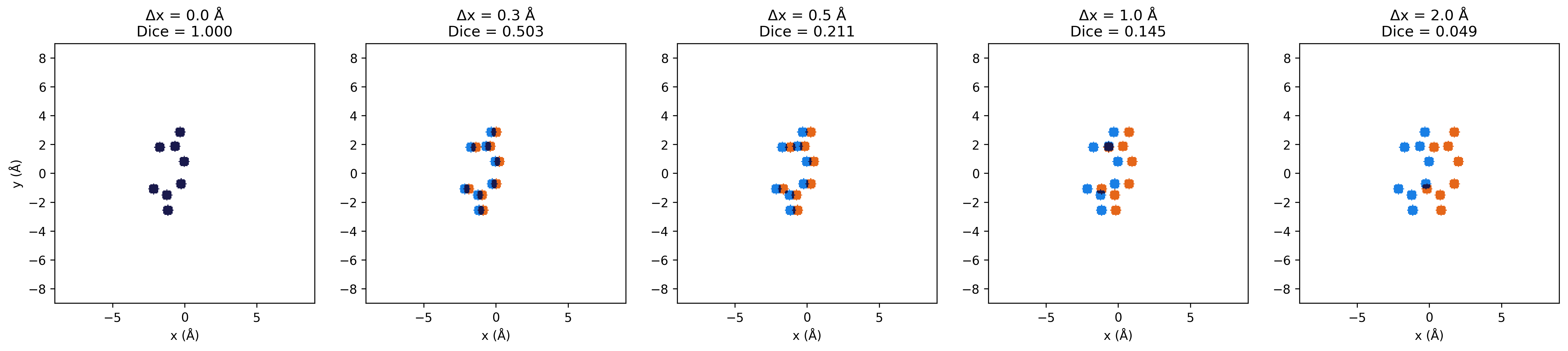}
    \caption{Sensitivity of the Dice similarity coefficient (DSC) to translational shifts. The DSC between a predicted atomic map and its ground truth drops rapidly as the prediction is shifted, showing that sub-\AA{} positional errors are heavily penalized.}
    \label{fig:dice_translation}
\end{figure}

In our setting, atoms are represented as small, sparse circles on a largely empty background. Although the DSC is a region-overlap measure, in this regime it responds sharply to positional error: Figure~\ref{fig:dice_translation} shows that translating a predicted map relative to the ground truth by even a fraction of an \AA{} produces a rapid drop in DSC. A misplaced atom therefore lowers the score substantially rather than being tolerated, so a high DSC reflects accurate localization and not merely a correct atom count. 

\subsection{Coordinate-Based Metrics}
\label{sec:coordinate_metrics}

While the Dice similarity coefficient (Eq.~\eqref{eq:dice}) provides a single, threshold-free measure of pixel-wise overlap, it does not on its own disentangle different aspects of structural accuracy, such as whether atoms are missed, spuriously detected, or merely displaced. To provide a more detailed picture of model performance, we additionally extract discrete atomic coordinates from the predicted probability maps using a Laplacian-of-Gaussian (LoG) blob-detection algorithm \cite{lindeberg1998feature, marr1980theory}, allowing us to report atom-count error, precision, recall, and coordinate RMSD.

For the atomic structure prediction model, at a representative binarization threshold of 0.2 this yields a mean atom-count error of 0.38, precision and recall of 0.98, and an atomic coordinate root mean square deviation (rmsd) of 0.097~\AA over correctly detected atoms. The sub-0.1~\AA{} localization error confirms that the model recovers atomic positions accurately.

\begin{figure}[h]
    \centering
    \includegraphics[width=\linewidth]{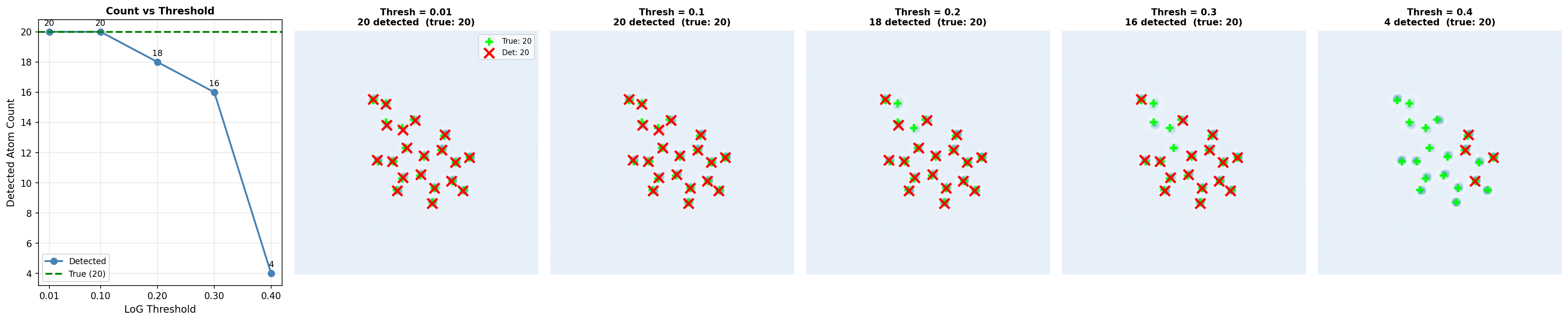}
    \caption{Sensitivity of the predicted atom count to the binarization threshold used in Laplacian of Gaussian (LoG) blob extraction.}
    \label{fig:threshold_sensitivity}
\end{figure}

\begin{figure}[h]
    \centering
    \includegraphics[width=\linewidth]{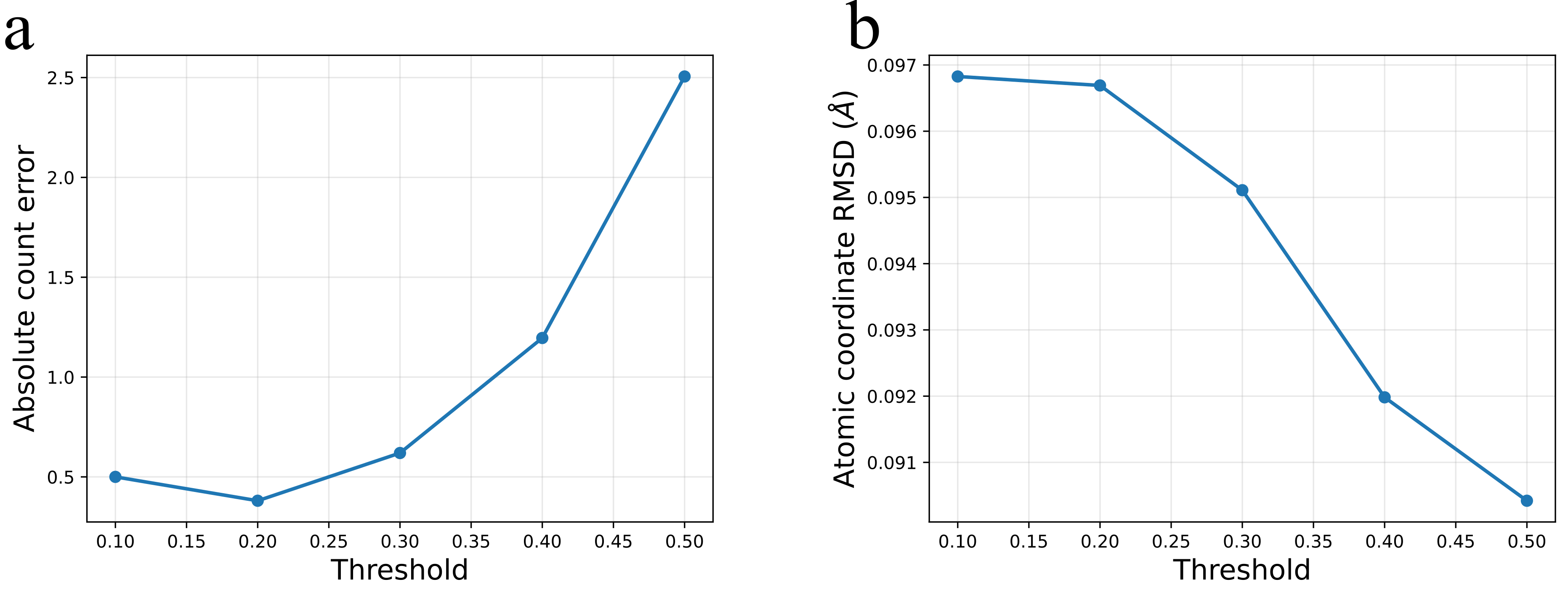}
    \caption{Dependence of coordinate based evaluation metrics on the LoG binarization threshold. (a) Mean absolute atom-count error vs. threshold. (b) Atomic coordinate RMSD vs. threshold.}
    \label{fig:extra_metrics}
\end{figure}

Extracting discrete coordinates in this way, however, requires a binarization threshold whose optimal value is not guaranteed to generalize across molecules with differing signal strengths. As shown in Fig.~\ref{fig:threshold_sensitivity}, the number of detected atoms varies considerably with the threshold, and Fig.~\ref{fig:extra_metrics} shows that both the atom-count error and the coordinate RMSD become volatile and artificially dependent on the chosen value. Relying on such threshold-dependent metrics would compromise the fully automated nature of the pipeline and obscure the network's raw, threshold-independent predictive capability. We therefore retain threshold-agnostic metrics such as the DSC as the primary measure of performance, and report these coordinate-level metrics as a complementary, corroborating view.

\begin{figure}
    \centering
    \includegraphics[width=\linewidth]{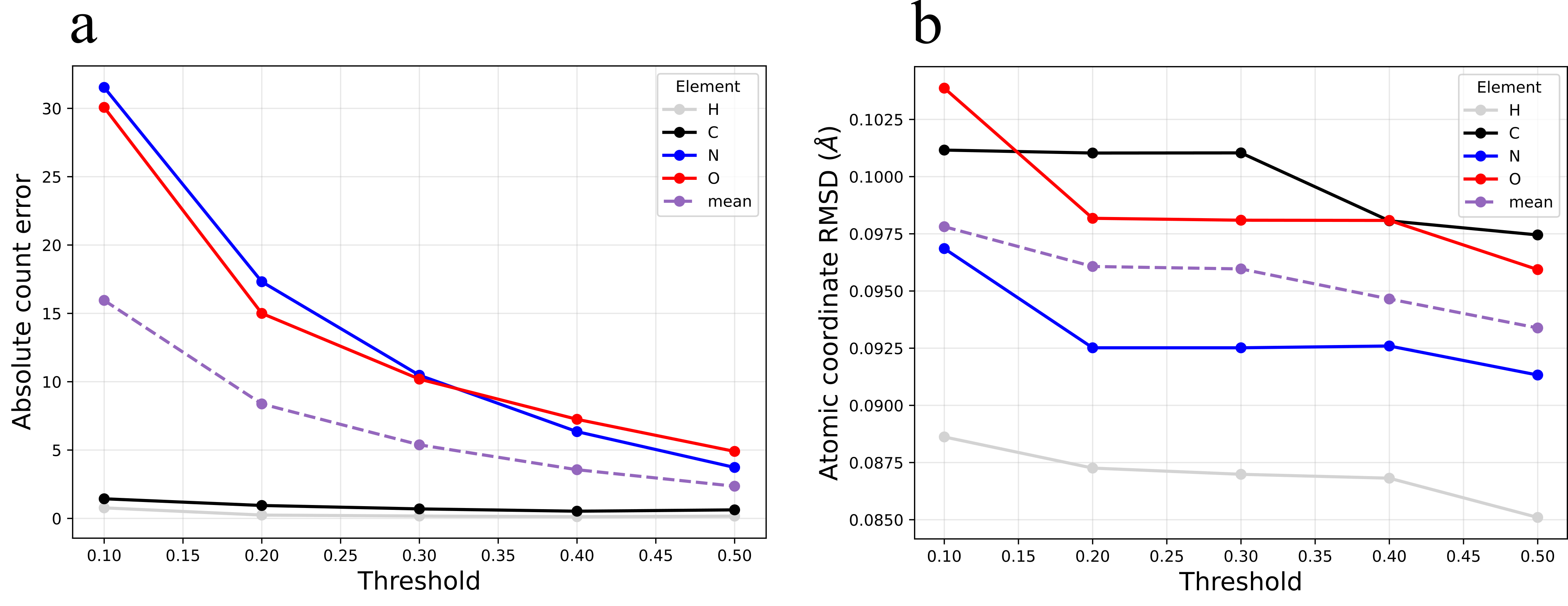}
    \caption{Dependence of the per-element coordinate based metrics for the chemically-resolved prediction model on the LoG binarization threshold. (a) Absolute atom-count error vs. threshold. (b) Atomic coordinate RMSD vs. threshold.}
    \label{fig:chem_threshold}
\end{figure}

The same analysis applied to the chemically-resolved prediction model is reported per element in Table~\ref{tab:metrics}, with the corresponding threshold dependence shown in Fig.~\ref{fig:chem_threshold}. For this model we use a binarization threshold of 0.5, which gave the best results across the values we tested (Fig.~\ref{fig:chem_threshold}). Hydrogen and carbon are detected and localized reliably (precision and recall $\approx 0.96$--0.98, atom-count error $<0.6$), whereas nitrogen and oxygen show substantially higher atom-count errors and lower precision and recall. This mirrors the DSC trend reported in Sec.~\ref{sec:chem_prediction} and reflects the same underlying cause, the underrepresentation of N and O in the training set, rather than a localization failure: when these atoms are detected they are placed accurately, as the coordinate RMSD remains uniformly low (${\sim}0.09$~\AA) across all four elements.

\begin{table}[ht]
  \centering
  \caption{Per-element detection and localization metrics for the chemically-resolved prediction model, obtained by LoG blob detection at a binarization threshold of 0.5.}
  \label{tab:metrics}
  \begin{tabular}{lccccc}
    \hline
    Element & Absolute count error & Precision & Recall & F1 & RMSD (\AA)  \\
    \hline
    H    & 0.158 & 0.983 & 0.974 & 0.976 & 0.085   \\
    C    & 0.560 & 0.971 & 0.960 & 0.961 & 0.097   \\
    N    & 3.663 & 0.663 & 0.657 & 0.649 & 0.091   \\
    O    & 4.554 & 0.696 & 0.682 & 0.685 & 0.096   \\
    \hline
    Mean & 2.234 & 0.828 & 0.818 & 0.818 & 0.092   \\
    \hline
  \end{tabular}
\end{table}

\section{Experimental dataset}
\label{sec:experimental}

\begin{figure}
    \centering
    \includegraphics[width=\linewidth]{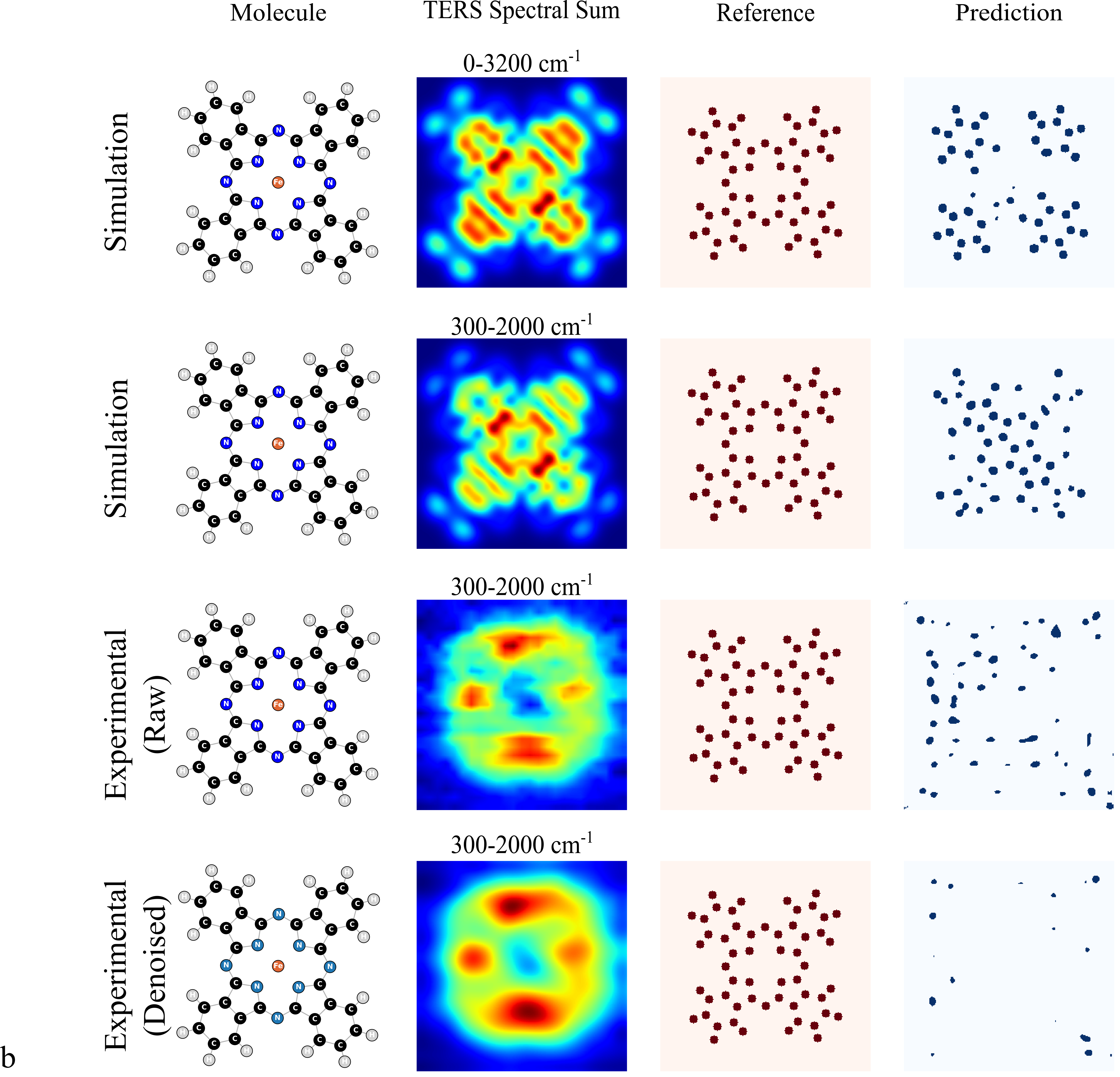}
    \caption{Comparison of geometry predictions for FePc on Ag(110) using simulated and experimental TERS data.}
    \label{fig:experimental_comparison}
\end{figure}

We performed an ablation study of our geometry prediction model using both experimental and simulated data for the phthalocyanine molecule with a Fe center (FePc). The experimental dataset used in our geometry prediction has been already published in Ref. \cite{Ferreira_experimental}, corresponding to TERS images of the FePc adsorbed on a Ag(110) surface in an aligned fashion. The authors kindly provided us with .stp files of the TERS hyperspectral maps taken at high intensity vibrations ranging from 300 to 2000 $\mathrm{cm}^{-1}$, whereas the simulated data span a broad range of 0 to 3200 $\mathrm{cm}^{-1}$, the full range of FePc vibrational modes.. Figure~\ref{fig:experimental_comparison} shows the geometry prediction on the simulated data, as well as two attempts on the experimental data: on the raw experimental data and after applying a gaussian function to denoise the image.

Regarding the simulated data, our model successfully predicts the periphery of the phthalocyanine molecule when the broad range of vibrational modes is used, although the predictions at the center of the molecule deviate noticeably from the ground truth. This behavior is already expected since Fe atom is not present in our dataset, limiting the capability of our model. Restricting the amount of simulated TERS images to the same frequency range of the experiment drastically deviates the geometry prediction from the ground truth, which is also expected since it does not correspond to the range in which the model was trained on.

Regarding the experimental data, the Figure \ref{fig:experimental_comparison} already shows a large mismatch between the simulated and experimental TERS spectral sum. Such mismatch is attributed to the simplifications made to acquire the simulation data, as opposed to the complex experimental environment faced by the FePc molecule. For example, our model considers a relative sharp tip (5 Å FWHM gaussian function) that is more appropriate for structural prediction at the Å level, whereas a blunter tip was most likely used in such experimental scenario. Additionally, it is also unknown how heavy is the influence of tip changes during experiment acquisition in the TERS images. Secondarily, substrate effects also drastically change some features in the TERS images, which are not present in our current model. Because of those discrepancies, the geometry prediction on the raw experimental data consists of several incongruent atomic positions that do not resemble the structure of the phthalocyanine molecule, and applying a denoise function drastically reduces the number of incongruent atomic positions, placing only a few at random locations.  

Given the remarks highlighted in the previous paragraphs, it is natural to expect that the model fails for the experimental images, especially given its limitations compared to the experiment. While our work still provides a proof-of-concept that TERS maps can be used for structural prediction, more robust models are necessary to achieve a better performance on experimental images. 

